\documentclass[12pt,
]{article}
\usepackage{amsmath, amsthm,mathtools, amssymb,upgreek,slashed,hyperref}
\usepackage[utf8]{inputenc}
\usepackage{ifpdf}
\ifpdf
  \usepackage[pdftex]{graphicx}
  \usepackage{xcolor}
  \usepackage{epstopdf}
  \usepackage{bigints}
\else
  \usepackage[dvips]{graphicx}
\fi
\usepackage[paper=letterpaper,margin=0.9in]{geometry}
\usepackage{tikz}
\def\Tr{{\rm Tr}}

\def\2{{\bf 2}}
\def\3{{\bf 3}}

\def\4{{\bf 4}}

\def\be{\begin{equation}}
\def\ee{\end{equation}}

\def\bar{\overline}

\def\tilde{\widetilde}

\def\hat{\widehat}

\font\teneurm=eurm10 \font\seveneurm=eurm7 \font\fiveeurm=eurm5
\newfam\eurmfam
\textfont\eurmfam=\teneurm \scriptfont\eurmfam=\seveneurm
\scriptscriptfont\eurmfam=\fiveeurm

\font\teneusm=eusm10 \font\seveneusm=eusm7 \font\fiveeusm=eusm5
\newfam\eusmfam
\textfont\eusmfam=\teneusm \scriptfont\eusmfam=\seveneusm
\scriptscriptfont\eusmfam=\fiveeusm

\font\tencmmib=cmmib10 \skewchar\tencmmib='177
\font\sevencmmib=cmmib7 \skewchar\sevencmmib='177
\font\fivecmmib=cmmib5 \skewchar\fivecmmib='177
\newfam\cmmibfam
\textfont\cmmibfam=\tencmmib \scriptfont\cmmibfam=\sevencmmib
\scriptscriptfont\cmmibfam=\fivecmmib

\numberwithin{equation}{section}

\def\d{\mathrm d}

\def\bar{\overline}

\def\vol{{\mathrm{vol}}}

\def\AdS{{\mathrm{AdS}}}

\begin{document}
\begin{titlepage}

\begin{center}

\vskip .5in 
\noindent

{\Large \bf{$\text{AdS}_2\times \text{S}^2\times \text{CY}_2$ solutions in Type IIB with 8 supersymmetries} }
\bigskip\medskip

Yolanda Lozano$^{a,}$\footnote{ylozano@uniovi.es}, Carlos Nunez$^{b,}$\footnote{c.nunez@swansea.ac.uk} and Anayeli Ramirez$^{a,}$\footnote{anayelam@gmail.com} \\

\bigskip\medskip
{\small

 $a$: Department of Physics, University of Oviedo,
Avda. Federico Garcia Lorca s/n, 33007 Oviedo, Spain
\vskip 3mm
 $b$: Department of Physics, Swansea University, Swansea SA2 8PP, United Kingdom}

\vskip .5cm 
\vskip .9cm 
     	{\bf Abstract }

\vskip .1in
\end{center}

\noindent
We present a new infinite family of Type IIB supergravity solutions preserving eight supercharges. The structure of the space is $\text{AdS}_2\times \text{S}^2\times \text{CY}_2\times \text{S}^1$ fibered over an interval. These solutions  can be related through double analytical continuations with those recently constructed in \cite{Lozano:2020txg}. Both types of solutions are however dual to very different superconformal quantum mechanics.  We show that our solutions fit locally in the class of   $\text{AdS}_2\times \text{S}^2\times \text{CY}_2$ solutions fibered over a 2d Riemann surface $\Sigma$ constructed by Chiodaroli, Gutperle and Krym, in the absence of D3 and D7 brane sources. We compare our solutions to the global solutions constructed by Chiodaroli, D'Hoker and Gutperle for $\Sigma$ an annulus. We also construct a cohomogeneity-two family of solutions using non-Abelian T-duality. Finally, we relate the holographic central charge of our one dimensional system to a combination of electric and magnetic fluxes. We propose an extremisation principle for the central charge from a functional constructed out of the RR fluxes. 

\noindent
\vskip .5cm
\vskip .5cm
\vfill
\eject

\end{titlepage}

\setcounter{footnote}{0}
\setcounter{tocdepth}{2}

{\parskip = .2\baselineskip \tableofcontents}

\section{Introduction}

\paragraph{}The Maldacena conjecture \cite{Maldacena:1997re} and its extensions motivate the search for AdS backgrounds preserving different amounts of Supersymmetry (SUSY), in different dimensions.

The half-maximal SUSY case is especially fructiferous. The correspondence between linear quiver conformal field theories preserving half-maximal SUSY and half-maximal BPS solutions with an AdS factor, leads to a precise map between infinite families of string backgrounds and their dual super-conformal field theories (SCFTs). Indeed, various works have developed the dictionary between $d$-dimensional SCFTs, the associated Hanany-Witten  brane set-ups \cite{Hanany:1996ie} and the dual AdS$_{d+1}$ string theory backgrounds.

For the case $d=6$, for which the strongly coupled conformal point is reached at high energies, the papers \cite{Apruzzi:2013yva, Gaiotto:2014lca, Cremonesi:2015bld, Nunez:2018ags}  have outlined the holographic dictionary and many other works have developed it. For $d=5$, the works  \cite{Brandhuber:1999np,Bergman:2012kr,Lozano:2012au,DHoker:2016ujz,DHoker:2016ysh,DHoker:2017zwj,Lozano:2018pcp} presented backgrounds with an AdS$_6$ factor and their UV-dual SCFTs. The dictionary for the case of  four dimensional ${\cal N}=2$ linear quiver SCFTs and their AdS$_5$ dual backgrounds was studied in \cite{Gaiotto:2009gz, ReidEdwards:2010qs, Aharony:2012tz, Nunez:2019gbg} among other works. The case of $d=3$ SCFTs (arising at low energies after a RG flow) and the dual AdS$_4$ backgrounds is studied in \cite{DHoker:2007zhm, DHoker:2008lup, Assel:2011xz} among other works. The correspondence for the case of two-dimensional (half-maximal BPS) low-energy SCFTs is particularly rich and has received a lot of attention recently. With the lens described above (linear quivers, Hanany-Witten set-ups and dual backgrounds), we encounter the works \cite{Couzens:2017way,Lozano:2019emq,Lozano:2019jza,Lozano:2019zvg,Lozano:2019ywa,Lozano:2020bxo,Faedo:2020nol,Faedo:2020lyw,Dibitetto:2020bsh} among various other papers.

The study of AdS$_2$ backgrounds in string/M-theory has a long and illustrious history. With the point of view described above, partial aspects of  the correspondence between super-conformal quantum mechanics theories (SCQMs) of the quiver type and half-maximal BPS backgrounds containing an AdS$_2$ factor were initially studied in \cite{Dibitetto:2018gtk, Gauntlett:2006ns, Kim:2013xza, Chiodaroli:2009yw, Chiodaroli:2009xh, Corbino:2018fwb, Corbino:2020lzq}. 
The recent works \cite{Dibitetto:2019nyz, Lozano:2020txg, Lozano:2020sae}
made precise and concrete the viewpoint advertised above for different infinite families of string backgrounds containing an AdS$_2$ factor.

This work presents a new infinite family of backgrounds with an AdS$_2$ factor.  We focus our presentation mostly on geometrical aspects of the new type IIB solutions. The contents of this paper are distributed as follows.
In Section \ref{B-type}, we present the new backgrounds preserving eight supersymmetries (four Poincar\'e and four conformal SUSYs). We study the conserved brane charges and 
deduce the associated brane set-up, consisting on D1 and D5 'colour' branes (dissolved into fluxes) with D3  and D7 'source' branes (present in the background and violating Bianchi identities). NS-five branes and fundamental strings complete this configuration. We define the holographic central charge following the procedure and physical meaning advanced  in  \cite{Lozano:2020txg}. The section is closed with a brief discussion of the dual SCQM. In Section \ref{CGKsection} we connect our backgrounds with those presented in \cite{Chiodaroli:2009yw, Chiodaroli:2009xh}. We point out  that the presence of sources in our solutions extend (for the AdS$_2$ fixed point) the results of \cite{Chiodaroli:2009yw, Chiodaroli:2009xh}. We also link the solutions in  \cite{Lozano:2020txg} with those of \cite{Chiodaroli:2009yw, Chiodaroli:2009xh} (under the above mentioned restrictions). These links require a zoom-in procedure that we discuss in detail.
In Section \ref{NATD} we uncover a new and explicit infinite family of solutions of cohomogeneity-two, by applying non-Abelian T-duality on the AdS$_3$ backgrounds of \cite{Lozano:2019emq, Lozano:2019jza, Lozano:2019zvg, Lozano:2019ywa}. The study of these backgrounds and their 'completion' following the ideas of \cite{Lozano:2016kum, Lozano:2016wrs, Lozano:2017ole, Itsios:2017cew, Lozano:2018pcp} 
is reserved for a future study.  We extend to the families of backgrounds discussed in this work a relation uncovered in  \cite{Lozano:2020txg},\cite{Lozano:2020sae} between the Ramond-Ramond sector of the backgrounds and the holographic central charge.  Such relation is discussed In Section \ref{maxwellcentralhol}. A functional whose extremisation yields the central charge is also presented.
Finally, Section \ref{conclusions} gives a short summary of the work, together with some ideas to work on the future.

\section{New AdS$_2\times \text{S}^2\times \text{CY}_2$ backgrounds}\label{B-type}

\paragraph{}In this section we present a new family of AdS$_2$ solutions with $\mathcal{N}=4$ Poincar\'e supersymmetry in Type IIB supergravity. These geometries are foliations of $\text{AdS}_2\times$S$^2\times$CY$_2\times$S$^1$   over an interval. Alternatively, they can be considered as foliations of  $\text{AdS}_2\times$S$^2\times$CY$_2$ over a 2d Riemann surface $\Sigma$ with the topology of an annulus.
The NS-NS sector of our solutions reads,

\begin{equation}\label{NS sector-B}
\begin{split}
\text{d}s_{st}^2 &= \frac{u \sqrt{\hat{h}_4 h_8} }{4 \hat{h}_4 h_8-(u')^2} \text{d}s^2_{\text{AdS}_2} + \frac{u}{4\sqrt{\hat{h}_4 h_8}} \text{d}s^2_{\text{S}^2}  + \sqrt{\frac{\hat{h}_4}{h_8}} \text{d}s^2_{\text{CY}_2} + \frac{\sqrt{\hat{h}_4 h_8}}{u} (\text{d} \psi^2+\text{d} \rho^2  )\, , \\
e^{- 2 \phi}&= \frac{h_8}{4\hat{h}_4} \big(4 \hat{h}_4 h_8 - (u')^2 \big) \, , \\ H_{3} &=- \frac{1}{2} \text{d} \bigg( \rho + \frac{u u'}{4 \hat{h}_4 h_8 - (u')^2} \bigg) \wedge  \text{vol}_{\text{AdS$_2$}}+ \frac{1}{h_8^2} \text{d} \rho \wedge H_2+\frac{1}{2} \text{vol}_{\text{S$^2$}} \wedge \text{d} \psi \, .
\end{split}
\end{equation}
Here $\phi$ is the dilaton, $H_3$ the NS-NS three-form and the metric is given in string frame. A prime denotes a derivative with respect to $\rho$. The two-form $H_2$ is defined on the $\text{CY}_2$.
The coordinate $\psi$ ranges in $[0, 2\pi]$, while the $\rho$ coordinate describes an interval that we will take to be bounded between $0$ and $2\pi(P+1)$ (see below). Note that  $u\ge 0$ and $4 \hat{h}_4 h_8 - (u')^2 \ge 0$ must be imposed to have a positive definite metric.
The background in eq.(\ref{NS sector-B}) is supported by the RR fluxes,
\begin{equation}\label{RR sector lower rank fluxes}
\begin{split}
F_{1} &= h_8' \text{d} \psi \, ,\\ F_{3} &=-  \frac{1}{h_8}H_2\wedge \text{d} \psi -\frac{1}{2} \Big( h_8 + \frac{h_8' u' u}{4 h_8 \hat{h}_4 - (u')^2} \Big) \text{vol}_{\text{AdS$_2$}} \wedge \text{d} \psi + \frac{1}{4} \left(- \text{d} \bigg(\frac{u'u}{2 \hat{h}_4} \bigg) + 2 h_8 \text{d} \rho \right) \wedge \text{vol}_{\text{S$^2$}}  \, , \\
F_{5} &= -(1+\star_{10})\left(\partial_{\rho} \hat{h}_4 \text{vol}_{\text{CY$_2$}} +\frac{h_8}{u} \hat{\star}_4 \text{d}_4 h_4 \wedge d \rho - \frac{u' u}{2h_8(4 \hat{h}_4 h_8 - (u')^2)} H_2 \wedge \text{vol}_{\text{AdS$_2$}} \right)\wedge\text{d}\psi.\\
F_{7} & =\left( \frac{1}{2}\left(\hat{h}_4+\frac{uu'\hat{h}_4'}{4 h_8 \hat{h}_4-(u')^2}\right)\text{vol}_{\text{AdS$_2$}}\wedge\text{d} \psi-\frac{1}{4}\left( 2\hat{h}_4\text{d}\rho-\text{d}\left(\frac{uu'}{2h_8}\right)\right)\wedge\text{vol}_{\text{S}^2}\right)\wedge\text{vol}_{\text{CY$_2$}} \, \\&-\star_{10}\left(\frac{1}{h_8}H_2\wedge \text{d} \psi\right), \\
F_{9} &=\frac{\hat{h}_4 h_8' u^2}{4 \hat{h}_8 (4 \hat{h}_4 h_8 - (u')^2)} \text{vol}_{\text{AdS$_2$}} \wedge\text{vol}_{\text{CY$_2$}} \wedge \text{vol}_{\text{S$^2$}} \wedge \mathrm{d} \rho \, .
\end{split}
\end{equation}
Supersymmetry holds whenever,
\begin{eqnarray}
\label{susyC}
u''=0,~~~~ H_2+ \hat{\star}_4 H_2=0,	
\end{eqnarray}
where $\hat{\star}_4$ is the Hodge dual on CY$_2$. In turn, the Bianchi identities of the fluxes impose--away from localised sources--that,
\begin{eqnarray}
\begin{split}
h_8''=0\, , \qquad \mathrm{d}H_2=0, \qquad\frac{h_8}{u}\nabla_{\text{CY}_2}^2\hat{h}_4+\partial_\rho^2\hat{h}_4-\frac{1}{h_8^3}H_2\wedge H_2=0.
\end{split}
\end{eqnarray}
In what follows we will concentrate on backgrounds for which $H_2=0$ and $\hat{h}_4=\hat{h}_4(\rho)$. These backgrounds are supersymmetric solutions of the Type  IIB equations of motion if the warping functions satisfy (away from localised sources),
\begin{eqnarray}
\hat{h}_4''=0,\qquad h_8''=0,\qquad u''=0,
\end{eqnarray}
 which makes them linear functions of $\rho$.

We focus  on the solutions defined by the piecewise linear functions $\hat{h}_4, h_8$ considered in \cite{Lozano:2019jza,Lozano:2019zvg}. These are continuous functions with discontinuous derivatives. These imply discontinuities in the RR-sector that are interpreted as generated by sources in the background. The solutions in  \cite{Lozano:2019jza,Lozano:2019zvg} have well-defined 2d dual CFTs. This requires a global definition of the $\rho$-interval. We achieve this imposing that $\hat{h}_4$ and $h_8$ vanish at both ends of the $\rho$-interval, that we take at $\rho=0,2\pi (P+1)$. Ending the space in this fashion, introduces extra source branes in the configuration. For the backgrounds to be trustable (in view of holographic applications), we need to impose that the sources are 'sparse', namely that they occur separated enough in the $\rho$-interval. This  imposes that $P$ (the length of the $\rho$-interval) is large.

The functions $\hat{h}_4$ and $h_8$  are then defined as,  
\begin{eqnarray} \label{profileh4sp}
\hat{h}_4(\rho)\!=\!\Upsilon\! \,h_4(\rho)\!=\!\!
                    \Upsilon\;\!\!\left\{ \begin{array}{cccrcl}
                       \frac{\beta_0 }{2\pi}
                       \rho & 0\leq \rho\leq 2\pi, &\\
                                     \alpha_k\! +\! \frac{\beta_k}{2\pi}(\rho-2\pi k) &~~ 2\pi k\leq \rho \leq 2\pi(k+1),& ~~k=1,...,P-1\\
                      \alpha_P-  \frac{\alpha_P}{2\pi}(\rho-2\pi P) &~~ 2\pi P\leq \rho \leq 2\pi(P+1),&
                                             \end{array}
\right.\\
\label{profileh8sp}
h_8(\rho)
                    =\left\{ \begin{array}{cccrcl}
                       \frac{\nu_0 }{2\pi}
                       \rho & 0\leq \rho\leq 2\pi, &\\
                        \mu_k+ \frac{\nu_k}{2\pi}(\rho-2\pi k) &~~ 2\pi k\leq \rho \leq 2\pi(k+1),& ~~k=1,...,P-1\\
                      \mu_P-  \frac{\mu_P}{2\pi}(\rho-2\pi P) &~~ 2\pi P\leq \rho \leq 2\pi(P+1).&
                                             \end{array}
\right.
\end{eqnarray}
The choice of constants is imposed by continuity of the metric and dilaton. This implies that
\begin{equation}
\alpha_k=\sum_{j=0}^{k-1} \beta_j, \qquad \mu_k=\sum_{j=0}^{k-1} \nu_j.
\end{equation}
In turn, $\beta_k$ and $\nu_k$ must be integer numbers to give well defined quantised charges (see the next subsection).  
In (\ref{profileh4sp}) the number $\Upsilon$  is chosen such that,
\begin{equation}
\Upsilon  \text{Vol}_{\text{CY$_2$}}= 16\pi^4.\label{upsilonnorm}
\end{equation}

In most of our analysis in this paper we will concentrate on solutions for which $u=u_0=$ constant. In that case the behaviour of the metric and dilaton at both ends of the $\rho$-interval is
\begin{equation}\label{O3O7brane-singularity}
\begin{split}
\text{d}s^2 \sim \frac{1}{x}(\text{d}s^2_{\text{AdS}_2} + \text{d}s^2_{\text{S}^2})  + \; \text{d}s^2_{\text{CY}_2} + x\; (\text{d} x^2+\text{d} \psi^2  )\, , \qquad
e^{-\phi}\sim x\,  ,
\end{split}
\end{equation}
where $x=\rho$ close to $\rho=0$ and $x=2\pi (P+1)-\rho$ close to $\rho=2\pi (P+1)$.
This corresponds to a superposition of D3-branes, extended on AdS$_2\times$S$^2$ and smeared on $\psi$ and the $\text{CY}_2$, and D7-branes, extended on AdS$_2\times$S$^2\times$CY$_2$ and smeared on $\psi$.  

\begin{figure}[h]
\centering
\includegraphics[scale=0.73]{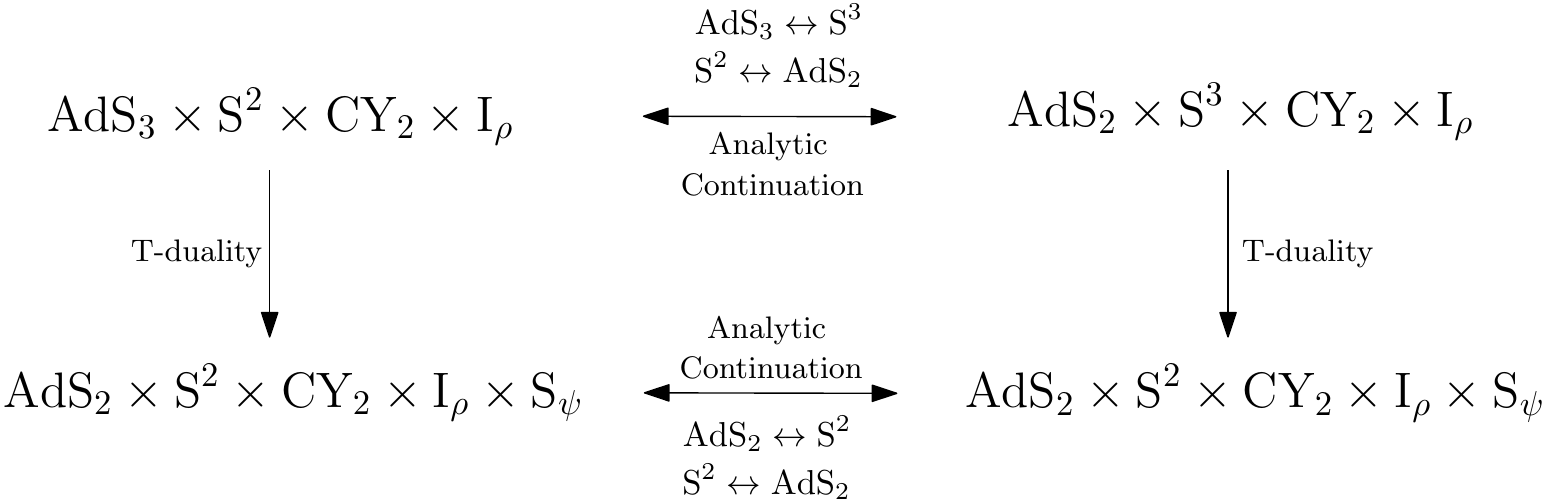}
\caption{Relations between the infinite family of AdS$_3$ backgrounds to massive IIA constructed in \cite{Lozano:2019emq} (top left), the IIB AdS$_2$ backgrounds studied in  \cite{Lozano:2020txg} (bottom left), the IIA AdS$_2$ backgrounds constructed in \cite{Lozano:2020sae} (top right), and the new AdS$_2$ solutions in Type IIB given by eqs.(\ref{NS sector-B})-(\ref{RR sector lower rank fluxes}) (bottom right).}
\label{diagrama}
\end{figure}

The backgrounds in eqs.\eqref{NS sector-B}-\eqref{RR sector lower rank fluxes} can be obtained applying the usual T duality rules  over the Hopf fibre  of the three sphere of the $\text{AdS}_2\times$S$^3$  backgrounds in \cite{Lozano:2020sae}. Additionally, these solutions have the same  structure as the geometries in \cite{Lozano:2020txg}, namely $\text{AdS}_2\times$S$^2\times$CY$_2\times$S$^1$ foliated over an interval. The relation with the backgrounds in  \cite{Lozano:2020txg}  is through an analytic continuation,
\begin{equation}
\begin{split}
&\text{d}s_{\text{AdS}_2}^2\to-\text{d}s_{\text{S}^2}^2	,\qquad \text{d}s_{\text{S}^2}^2\to-\text{d}s_{\text{AdS}_2}^2, \qquad e^{\phi}\to ie^{\phi},\qquad F_{i}\to -iF_{i},\\
&u\to -iu,\qquad \hat{h}_4\to i\hat{h}_4, \qquad h_8\to ih_8, \qquad\rho\to i\rho,\qquad\psi\to-i\psi, \qquad g_{i}\to i g_{i}.\label{diagramaeq}
\end{split}
\end{equation}
These relations are summarised in Figure \ref{diagrama}.

Next we  study the charges  associated with the backgrounds in eqs. \eqref{NS sector-B}-\eqref{RR sector lower rank fluxes}  and the  associated brane set-up.  
\subsection{Brane charges and brane set-up}
\paragraph{}
We compute the charges associated to our backgrounds using that the magnetic charge for a Dp brane is given by,
\begin{eqnarray}\label{chargemagnetic}
Q_{\text{Dp}}^{m}
=\frac{1}{(2\pi)^{7-p} }\int_{\mathcal{M}_{8-p}} \hat{F}_{8-p},\end{eqnarray}
where $\mathcal{M}_{8-p}$ is any $(8-p)$-dimensional compact manifold transverse to the branes.
In turn, the electric charge of Dp branes is defined by,
\begin{eqnarray}
Q_{\text{Dp}}^e=\frac{1}{(2\pi)^{p+1}}\int_{\text{AdS}_2\times \Sigma_{p}} \hat{F}_{p+2},\label{electric-charges}
\end{eqnarray}
where $\Sigma_p$ is the $p$-dimensional manifold on which the brane extends.
In both expressions we have set $\alpha'= g_s = 1 $. For the electric charges we need to regularise the volume of the AdS$_2$ space. We take it to be the analytical continuation of the volume of the two-sphere,  
\begin{equation} \label{regpres}
\text{Vol}_{\text{AdS}_2}=4\pi\, .
\end{equation}
In the previous expressions $\hat{F}$ are the Page fluxes, defined as $\hat{F}=F\wedge e^{-B_{2}}$.
They read, for our backgrounds
\begin{equation}\label{fluxes23}
\begin{split}
\hat{F}_{1} &= h'_8 \, \mathrm{d} \psi \, , \\
\hat{F}_{3} &= \frac{1}{2} \left(h'_8(\rho-2\pi k)  - h_8\right) \, \text{vol}_{\text{AdS$_2$}} \wedge \mathrm{d} \psi  + \frac{1}{4} \bigg(2h_8+\frac{u'(u\hat{h}_4'-\hat{h}_4u')}{2\hat{h}_4^2} \bigg)\text{vol}_{\text{S$^2$}}\wedge \text{d}\rho  \, , \\
\hat{F}_{5} &=\frac{1}{4}\left(h_8(\rho-2\pi k)- \frac{\left(u - (\rho-2\pi k)  u' \right) ( u  \hat{h}_4' - \hat{h}_4  u'  )}{4 \hat{h}_4^2}\right)
\text{vol}_{\text{AdS$_2$}} \wedge \text{vol}_{\text{S$^2$}} \wedge \mathrm{d} \rho - \hat{h}_4' \text{vol}_{\text{CY$_2$}} \wedge \mathrm{d} \psi \, , \\
\hat{F}_{7} &= \frac{1}{2} (\hat{h}_4 - (\rho-2\pi k) \hat{h}_4')\;  \text{vol}_{\text{AdS$_2$}}  \wedge \text{vol}_{\text{CY$_2$}} \wedge \mathrm{d} \psi - \frac{1}{4}\left( 2\hat{h}_4+\frac{u'(uh_8'-h_8u')}{2h_8^2} \right) \text{vol}_{\text{S$^2$}} \wedge \text{vol}_{\text{CY$_2$}}\wedge\text{d}\rho \, ,\\
\hat{F}_{9}&= -\frac{1}{4}\left((\rho-2\pi k) \hat{h}_4 -\frac{\left(u - (\rho-2\pi k)  u' \right) ( u  h_8' - h_8 u')}{4h_8^2}\right)
\text{vol}_{\text{AdS$_2$}} \wedge \text{vol}_{\text{S$^2$}}\wedge  \text{vol}_{\text{CY$_2$}}\wedge d\rho.
\end{split}
\end{equation}
In these expressions we have allowed for large gauge transformations of the $B_2$-field, $B_2\to B_2+ k \pi \text{vol}_{\text {AdS$_2$}}$, as in  \cite{Lozano:2020sae} (see this reference for more details).

Before calculating the quantised charges associated to these fluxes it is useful to compute the following quantities,
\begin{equation}\label{caxa2}
\begin{split}
\text{d}\hat{F}_{1} &= h_8'' \text{d}\rho\wedge \text{d}\psi \,, \quad
\text{d}\hat{F}_{3} = \frac{1}{2}h_8'' (\rho-2\pi k)    \text{vol}_{\text{AdS$_2$}} \wedge\mathrm{d} \rho\wedge \mathrm{d} \psi\, , \quad
\text{d}\hat{F}_{5} = -h_4'' \; \text{vol}_{\text{CY$_2$}} \wedge \mathrm{d}\rho\wedge\mathrm{d} \psi \, , \\
\text{d}\hat{F}_{7} &= -\frac{1}{2}h_4'' (\rho-2\pi k)  \text{vol}_{\text{AdS$_2$}}  \wedge \text{vol}_{\text{CY$_2$}} \wedge\mathrm{d}\rho \wedge \mathrm{d} \psi\, , \quad
\text{d}\hat{F}_{9} = 0 \, .
\end{split}
\end{equation}
In these expressions $\hat{h}_4''$ and $h_8''$ are the ones that follow from eqs.\eqref{profileh4sp}-\eqref{profileh8sp},
\begin{eqnarray}
\label{nana}
& & \hat{h}_4''=\frac{1}{2\pi}\sum_{k=1}^P (\beta_{k-1}-\beta_{k}) \delta(\rho-2\pi k),\;\;\;\; h_8''=\frac{1}{2\pi}\sum_{k=1}^P (\nu_{k-1}-\nu_{k} )\delta(\rho-2\pi k),\\
& & \hat{h}_4'' \times(\rho-2\pi k) = h_8'' \times(\rho-2\pi k) =x \delta(x)=0.\nonumber
\end{eqnarray}
We then obtain
\begin{equation}
\text{d}\hat{F}_{3}=\text{d}\hat{F}_{7}=0,
\end{equation}
and
\begin{eqnarray}
\text{d}\hat{F}_{1} &=&\frac{1}{2\pi}\sum_{k=1}^P (\nu_{k-1}-\nu_{k} )\delta(\rho-2\pi k) \,\text{d}\rho\wedge \text{d}\psi \\
\text{d}\hat{F}_{5} &=& -\frac{1}{2\pi}\sum_{k=1}^P (\beta_{k-1}-\beta_{k}) \delta(\rho-2\pi k)\, \text{vol}_{\text{CY$_2$}} \wedge \mathrm{d}\rho\wedge\mathrm{d} \psi.
\end{eqnarray}
These results can be put in correspondence with the brane set-up summarised in Table \ref{Table brane web set type IIA2}. 
\begin{table}[ht]
\begin{center}
\begin{tabular}{|c|c|c|c|c|c|c|c|c|c|c|}
	\hline	
& $x^0=t$ & $x^1$ & $x^2$ & $x^3$ & $x^4$ & $x^5=\rho$ & $x^6=r$ & $x^7=\theta_1$ & $x^8=\theta_2$ & $x^9=\psi$\\
\hline
D1 & x & & & & & $$ & & &$$ &x  \\
\hline
 D3 &  x & &  &  &  & & x & x & x &  \\
\hline
D5 & x & x & x & x & x & $$ & & & &x \\
\hline
D7 & x & x & x & x & x & & x & x & x &  \\
\hline
NS5 & x & x & x & x & x & & x& & & $$ \\
\hline F1 &x & $$ & $$ & $$ & $$ & $$& x & & & $$ \\ 	\hline	
\end{tabular}
\caption{Brane set-up associated to our solutions. $x^0$ corresponds to the time direction of the ten dimensional spacetime, $x^1, \dots , x^4$ are the coordinates spanned by the CY$_2$ and $ x^7 , x^8$ are the coordinates parametrising the $S^2$.}
\label{Table brane web set type IIA2}
\end{center}
\end{table}
The fact that $\text{d}\hat{F}_{3}=0$ and $\text{d}\hat{F}_{7}=0$ indicates that the D5 and D1 branes play the role of colour branes (dissolved in fluxes) in the brane set-up. On the other hand, $\text{d}\hat{F}_{1}$ and $\text{d}\hat{F}_{5}$ being nonzero indicate that the D7 and D3 branes are flavour branes, that is, explicit sources with dynamics described by the Born-Infeld-Wess-Zumino action.

Substituting $\hat{h}_4$ and $h_8$ as defined by eqs.(\ref{profileh4sp}) and (\ref{profileh8sp}),  together with eqs.(\ref{upsilonnorm}) and (\ref{regpres}), we find, in each $\rho$-interval $[2\pi k,2\pi(k+1)]$
\begin{eqnarray}
\begin{split}
	 Q_{\text{D1}}^{e}&= 
	\frac{1}{(2\pi)^2}\int_{\text{AdS}_2\times\text{S}_\psi} \hat{F}_3=\left(\frac{ \text{Vol}_{\text{AdS$_2$} }}{4\pi} \right)\left(\frac{ \text{Vol}_{\psi}}{\pi} \right)\frac{h_8-h'_8(\rho-2\pi k)}{2}=\mu_k,\\
  Q_{\text{D3}}^{m}&=\frac{1}{16\pi^4}\int_{\text{CY}_2\times\text{S}_{\psi}} \hat{F}_5=  \frac{1}{16\pi^4}\int_{\text{CY}_2\times\text{S}_{\psi}\times\text{I}_\rho} \text{d}\hat{F}_5=\left(\frac{ \Upsilon \text{Vol}_{\text{CY$_2$} }}{16\pi^4} \right) \times \text{Vol}_\psi~ \int h_4''\text{d}\rho=(\beta_{k-1}-\beta_{k}),\\
 Q_{\text{D5}}^{e}&=\frac{1}{(2\pi)^6}\int_{\text{AdS}_2\times\text{CY}_2\times\text{S}_{\psi}} \hat{F}_7=\left(\frac{ \text{Vol}_{\text{AdS$_2$} }}{4\pi} \right)\left(\frac{ \Upsilon \text{Vol}_{\text{CY$_2$} }}{16\pi^4} \right)\left(\frac{ \text{Vol}_{\psi}}{\pi} \right)\frac{h_4-h'_4(\rho-2\pi k)}{2}=\alpha_k,\\
 Q_{\text{D7}}^{m}&=\int_{\text{S}_\psi} \hat{F}_1=\int_{\text{S}_{\psi}\times\text{I}_\rho} \text{d} \hat{F}_1=  \text{Vol}_\psi ~\int h_8'' \text{d}\rho=(\nu_{k-1}-\nu_{k}).
 \label{cargaspageB}
\end{split}	
\end{eqnarray}
Further, in the brane set-up the F1-strings are electrically charged with respect to the NS-NS 3-form $H_3$ while the NS5 branes are magnetically charged,   
\begin{eqnarray}
\begin{split}
	 Q_{\text{F1}}^{e}&=\frac{1}{(2\pi)^2}\int_{\text{AdS}_2\times\text{I}_\rho} H_3=\left(\frac{ \text{Vol}_{\text{AdS$_2$} }}{4\pi} \right)\left(\frac{1}{2\pi} \right)\int_{2\pi k}^{2\pi (k+1)} \text{d}\rho=1,\\
	 Q_{\text{NS5}}^{m}&=\frac{1}{(2\pi)^2}\int_{\text{S}^2\times\text{S}_{\psi}} H_3=\left(\frac{ \text{Vol}_{\text{S$^2$} }}{4\pi} \right)\left(\frac{ \text{Vol}_{\psi}}{2\pi} \right)=1 .
\end{split}
\end{eqnarray}

\subsection{Holographic central charge}\label{hccharge}

\paragraph{} To close this part of our study we compute the holographic central charge associated to our solutions. Being the field theory zero-dimensional, the previous quantity should be interpreted as the number of vacuum states in the dual  superconformal quantum mechanics (see  \cite{Lozano:2020txg, Lozano:2020sae} for a further discussion of the physical meaning of this quantity). We follow the prescription in \cite{Macpherson:2014eza, Bea:2015fja}.
We get for the internal volume,
\begin{equation}\label{Vint-Btype}
\begin{split}
V_{\text{int}} &= \int \text{d}^8x\sqrt{e^{-4\phi}\;\text{det}\;g_{8,ind}} = \frac{\text{Vol}_{\text{CY}_2} \text{Vol}_{\text{S}^2} \text{Vol}_{\psi}}{4^2} \int_0^{2\pi(P+1)}(4\hat{h}_4h_8-(u')^2)\;\text{d}\rho \, ,
\end{split}
\end{equation}
and, finally, for the central charge
\begin{equation}\label{hccharge-B}
c_{\text{hol,1d}}=\frac{3 V_{\text{int}}}{4\pi G_N}	=\frac{3}{\pi}\int_0^{2\pi(P+1)} \left({h}_4h_8 - \frac{(u')^2}{4\Upsilon} \right) \text{d} \rho \, .
\end{equation}
We have used that $G_N=8\pi^6$ and set units so that $\alpha'=g_s=1$. 

We would like to stress that in the usual calculations, such as the previous one, giving rise to the holographic central charge, only the NS-NS sector of the backgrounds needs to be taken into account. We will point out an interesting relation between the holographic central charge and the RR sector of our AdS$_2$ solutions in Section \ref{maxwellcentralhol}. Such relation has been previously encountered in the AdS$_2$ solutions constructed in \cite{Lozano:2020txg,Lozano:2020sae}. 
\subsection{Aspects of the dual Conformal Quantum Mechanics }
\paragraph{}
Whilst the main focus of this work in not the Quantum Mechanical analysis of the duals to the backgrounds in eqs.(\ref{NS sector-B})-(\ref{RR sector lower rank fluxes}), we add below some thoughts along this direction.

In the papers \cite{Lozano:2019jza,Lozano:2019zvg,Lozano:2020txg,Lozano:2020sae} concrete quivers were proposed
as UV-descriptions of weakly coupled 2d QFTs or 1d Quantum Mechanics. It was conjectured that these quivers become strongly coupled at low energies and a conformal fixed point arises. Checks for these proposals were presented in each of the works \cite{Lozano:2019jza,Lozano:2019zvg,Lozano:2020txg,Lozano:2020sae}, for the different systems under study. These checks deal with RG-invariant quantities that can be well-identified in the UV and IR descriptions.

As we indicated around eq.(\ref{diagramaeq}) and summarised in Figure \ref{diagrama}, the backgrounds of Section \ref{B-type} arise after an Abelian T-duality on the backgrounds of  \cite{Lozano:2020sae}.
This suggests that the quantum mechanical system proposed in \cite{Lozano:2020sae} should also apply here.
We are in fact T-dualising across a non-R-symmetry-direction, hence we expect the amount of SUSY to be the same. The R-symmetry of the quivers in  \cite{Lozano:2020sae} is $SU(2)_R$, and there is also 
a global $SU(2)_g$ symmetry. We are choosing a $U(1)_g$ inside $SU(2)_g$ for our dualisation. Therefore, our dual quantum mechanical system should have $SU(2)_R\times U(1)_g$ symmetry. This is in fact 
geometrically realised by the presence of the round S$^2$ and the circle S$^1_\psi$ in the backgrounds of Section \ref{B-type}.

Since the string sigma model in a background and in its T-dual is the same, we expect the same dual quantum mechanical systems 
for our backgrounds as those for the backgrounds \cite{Lozano:2020sae} (only that perhaps it will be written in a different language).

Using this reasoning, we may think about the SCQM as that arising in the very low energy limit of a system of D3-D7 branes---dual to a four dimensional ${\cal N}=2$ QFT. This system is 'polluted' by 
one-dimensional defects. These are Wilson loops (arising from F1-D5) and 't Hooft loops (arising from NS5-D1) added to the background, see for example \cite{Assel:2019iae}. Note that both the D1's and the D5's
extend on the $\psi$-isometric direction.
From the discussion above, it follows that the dual SCQM to our backgrounds is the description of these one-dimensional defects inside a four dimensional ${\cal N}=2$
 QFT. In fact, in the IR the gauge symmetry on both D7 and D3 branes should become global. This implies that these branes must be sources/flavours, as it occurs  in the backgrounds of Section \ref{B-type}.
 By the same token we have two lines of one dimensional gauge groups: $\Pi_{i=1}^{P} U(\alpha_i)$ and $\Pi_{i=1}^{P} U(\mu_i)$ realised on D5 and D1 branes in each $\rho$-interval. 
 This is reflected by the counting of branes of eq.(\ref{cargaspageB}). The nodes in the $[2\pi k, 2\pi(k+1)]$ interval will have $SU(\beta_k-\beta_{k+1})$ and $SU(\nu_k-\nu_{k+1})$ flavour groups, realised 
 on the D3 and D7 branes,
 as also reflected by eq.(\ref{cargaspageB}). The brane set-up  is the one described in Table \ref{Table brane web set type IIA2}.
 
 As was found in \cite{Lozano:2020sae}, our 1d system should also have Wilson lines (in an antisymmetric representation) inserted in the different gauge nodes of the quiver. These Wilson lines arise from the 
 massive fermionic strings that stretch between D1s in the $k$-th interval and D7s in all other intervals. The Wilson lines would be in the $(\nu_0,\dots, \nu_{k-1}$) antisymmetric representation of the $U(\mu_k)$ gauge group. The same applies to the massive D3-D5 fermionic strings and the antisymmetric Wilson lines 
 on the $U(\alpha_k)$ groups. As in \cite{Lozano:2020sae}, this information can be encoded in Young diagrams.
 
 We would also have a dynamical CS-term of each gauge group. This comes from the massless fermionic strings
 stretched  between D1-D7 and D5-D3 branes. The coefficient can be extracted studying the WZ action for a D1 along $[t,\psi]$ and a D5 along $[t, \text{CY}_2, \psi]$. As expected, these coefficients are quantised.
 
\begin{figure}[h]
\centering
\includegraphics[scale=0.55]{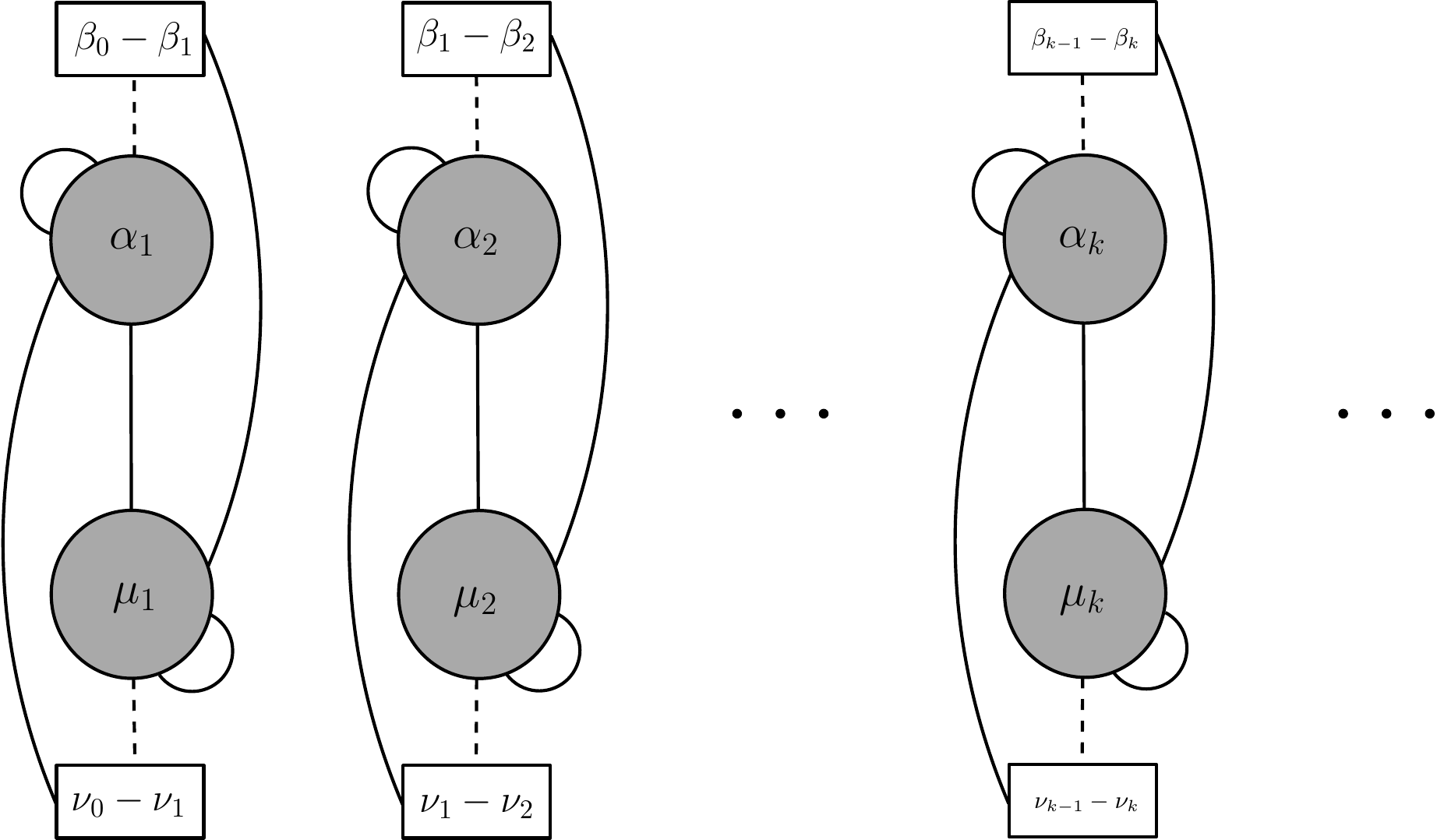}
\caption{The proposed quantum mechanical quiver. This follows from the analysis of open strings in the Hanany-Witten set-up.}
\label{quiverprop}
\end{figure}

 The field content of the UV-quantum mechanical quiver 
follows directly from the analysis of Appendix B in  \cite{Lozano:2020sae}. In fact, each node contains a $(4,4)$ vector multiplet and a $(4,4)$ adjoint hyper, $(0,4)$ bifundamental hypers join the two types of colour, D5 and D1, branes, (4,4) bifundamental hypers join the D7 sources with D5-branes, and the  
D3 sources with  D1 branes, respectively. Finally, $(0,2)$ Fermi multiplets join source D7 with colour D1s and source D3 with colour D5 branes. The quiver diagram is 
depicted in Figure \ref{quiverprop}.

\section{Connection with the $\text{AdS}_2\times \text{S}^2\times \text{CY}_2\times \Sigma$ backgrounds of Chiodaroli-Gutperle-Krym} \label{CGKsection}

\paragraph{} In this section we relate our backgrounds to the general class of  $\text{AdS}_2\times \text{S}^2\times \text{CY}_2\times \Sigma$ solutions to Type IIB supergravity with 8 supercharges found by Chiodaroli, Gutperle and Krym (CGK) in \cite{Chiodaroli:2009yw}. We show that our solutions fit locally in their classification in the absence of D3 and D7 brane sources (in this sense our backgrounds extend those in \cite{Chiodaroli:2009yw} at the AdS$_2$ point). A similar analysis shows that the family of $\text{AdS}_2$ solutions  to Type IIB supergravity recently found in \cite{Lozano:2020txg} also fits in their general class.

\subsection{Review of the CGK geometries}\label{Gutperlething}

\paragraph{} The CGK backgrounds are dual to one dimensional conformal interfaces inside the two dimensional CFT associated to the D1-D5 system. These solutions (unlike ours) interpolate between AdS$_2$ in the IR (at the interface) and the AdS$_3\times \text{S}^3\times \text{CY}_2$ solution of Type IIB supergravity in the UV. We shall focus on the AdS$_2$ fixed points and compare them with both the backgrounds discussed in section \ref{B-type} and  the solutions found  in  \cite{Lozano:2020txg}. 

In \cite{Chiodaroli:2009yw}, the authors used techniques developed in \cite{DHoker:2007zhm, DHoker:2007mci} to find half BPS solutions that preserve eight of the sixteen supersymmetries of the AdS$_3\times$S$^3\times$CY$_2$ vacuum, and are locally asymptotic to this vacuum solution.  They provided an ansatz for the bosonic fields in Type IIB supergravity for a  foliation of AdS$_2\times$S$^2\times $CY$_2$ over a two-dimensional Riemann surface $\Sigma$ with a boundary, and found that 
the local solutions of the BPS equations can be written in terms of two harmonic and two holomorphic functions defined on $\Sigma$.  The solutions corresponds to a D1-D5 configuration with extra NS5 branes and fundamental strings, but vanishing D3 and D7 brane charges. We will see that our solutions fit locally within this class of solutions in the absence of  D3 and D7  brane sources. Our D3 and D7 sources are localised in $\rho$ and smeared in the $\psi$-coordinate. The mapping explained below is valid at points in $\rho$  where the sources are not present.


We start summarising the local solutions constructed in \cite{Chiodaroli:2009yw}.  The metric for the ten-dimensional spacetime is given, in Einstein frame, by
\begin{equation}\label{mmaacc}
\text{d}s^2=f_1^2\text{d}s_{\text{AdS}_2}^2+f_2^2\text{d}s_{\text{S}^2}^2+f_3^2\text{d}s_{\text{CY}_2}^2+\tilde{\rho}^2\text{d}{z}\text{d}{\bar{z}} \, ,
\end{equation}
where the warping factors $f_i$ ($i=1, \dots, 3$) and $\tilde{\rho}$ are functions of $z$ and $\bar{z}$, the  local holomorphic coordinates of $\Sigma$. The orthonormal frames can be written as,
\begin{alignat}{2}
\begin{split}
   f_1^2 \text{d}s_{\text{AdS}_2}^2 = \eta_{i_1 i_2} e^{i_1} \otimes	e^{i_2} \, ,& \quad \quad \quad  i_{1,2} = 0, 1\, ,\\
   f_2^2 \text{d}s_{\text{S}^2}^2 = \delta_{j_1 j_2}e^{j_1} \otimes e^{j_2} \, ,&  \quad \quad \quad j_{1, 2} = 2, 3 \, , \\
   f_3^2 \text{d}s_{\text{CY}_2}^2 = \delta_{k_1 k_2} e^{k_1} \otimes	e^{k_2}  \, ,& \quad \quad \quad  k_{1, 2} = 4, 5, 6, 7\, , \\
   \tilde{\rho}^2 \text{d}{z} \text{d}{\bar{z}} = \delta_{a b}e^{a} \otimes e^{b}  \, ,& \quad \quad \quad  a, b = 8, 9\, .
   \end{split}
\end{alignat}
The NS-NS and RR  three-forms are written as a complex three-form, defined as $G=e^{\Phi}H_{3} + i e^{-\Phi}(F_{3} - \chi \;H_{3})$. This form is given by,
\begin{equation}\label{3-formComplex}
G=g^{(1)}_ae^{a01}+g^{(2)}_ae^{a23} \, .
\end{equation}
In turn, the self-dual five-form flux is,
 \begin{equation}\label{5-form}
 F_{5} = h_ae^{a0123} + \tilde{h}_ae^{a4567} \, , \quad  a = z, \bar{z} \, ,
\end{equation}
where the self-duality condition implies $h_a=-\epsilon_a\,^b\tilde{h}_b$.

The local solutions of the BPS equations and Bianchi identities admit a description in terms of four functions, $A$, $B$, $H$ and $K$. The analysis in \cite{Chiodaroli:2009yw} shows that the functions $A$ and $B$ must be holomorphic on the Riemann surface $\Sigma$, whilst $H$ and $K$ must be harmonic. The supergravity fields can be written in terms of these functions as,
\begin{equation}
\begin{split}
\label{mmbbx}
f_1^2&=\frac{e^{-2\Phi}|H|}{2f_3^2K}\left((A+\bar{A})K-(B-\bar{B})^2\right) \, , \quad\quad
f_2^2=\frac{e^{-2\Phi}|H|}{2f_3^2K}\left((A+\bar{A})K-(B+\bar{B})^2\right) \, , \\
f_3^4&=4\frac{e^{2\Phi}K}{A+\bar{A}} \, , \quad\quad
e^{4\Phi} = \frac{1}{4K^2}\left( ( A + \bar{A}) K - ( B + \bar{B})^2 \right) \left( ( A + \bar{A}) K - (B - \bar{B})^2 \right) \, ,\\
\chi &=\frac{1}{2 i K} \left( B^2 - \bar{B}^2 - ( A - \bar{A}) K \right) \, , \quad\quad \tilde{\rho}^4=e^{2\Phi}K\frac{(A+\bar{A})}{H^2}\frac{|\partial_zH|^4}{|B|^4} \, .
\end{split}
\end{equation}
Here $\Phi=-\phi/2$, where $\phi$ is the dilaton.
For the five-form field strength, we define a four-form potential, along CY$_2$, 
\begin{equation}\label{four-potential}
C_{\text{CY}_2} = -\frac{i}{2}\frac{ B^2 - \bar{B}^2}{ A + \bar{A}} - {2} \tilde{K}, \qquad \qquad \partial_z C_{\text{CY}_2} = f_3^4 \tilde{\rho}\;\tilde{h}_z \, ,
\end{equation}
where $\tilde{K}$ is the harmonic function conjugate\footnote{\label{conjugatefoot}The harmonic conjugate of $g$ is denoted as $\tilde{g}$ and satisfies $i\partial_z \tilde{g}=\partial_z g$.} to $K$.

The potentials for  the field strengths in equation \eqref{3-formComplex} are written in terms of the holomorphic and harmonic functions as
\begin{equation}
\begin{split}
\label{potential-b1}
b^{(1)}=-\frac{H(B+\bar{B})}{(A+\bar{A})K-(B+\bar{B})^2}-h_{1},\qquad &h_1=\frac{1}{2}\int\frac{\partial_zH}{B}+\text{c.c.}\;,\\
b^{(2)}=-i\frac{H(B-\bar{B})}{(A+\bar{A})K-(B-\bar{B})^2}+\tilde{h}_{1},\qquad &\tilde{h}_1=-\frac{i}{2}\int\frac{\partial_zH}{B}+\text{c.c.}\;,\\
c^{(1)}=-i\frac{H(A\bar{B}-\bar{A}B)}{(A+\bar{A})K-(B+\bar{B})^2}+\tilde{h}_{2},\qquad &\tilde{h}_2=-\frac{i}{2}\int\frac{A\;\partial_zH}{B}+\text{c.c.}\;,\\
c^{(2)}=-\frac{H(A\bar{B}+\bar{A}B)}{(A+\bar{A})K-(B-\bar{B})^2}+h_{2},\qquad &h_2=\frac{1}{2}\int\frac{A\;\partial_zH}{B}+\text{c.c.},\;
\end{split}
\end{equation}
where $\tilde{h}_i$ and $h_i$ are  harmonic functions conjugate to each other. In the previous expression $b^{(1)}$ and $b^{(2)}$ are the potentials of the NS-NS three-form $H_{3}$ and $c^{(1)}$ and $c^{(2)}$ are the potentials related to the RR three-form $F_{3}$. These read, 
\begin{equation}
\begin{split}
H_3=&\text{d}b^{(1)}\wedge \vol_{\AdS_2}+\text{d}b^{(2)}\wedge \vol_{\text{S}^2}\\
F_3=&\d C_2-\chi H_3=(\d c^{(1)}-\chi\text{d}b^{(1)})\wedge \vol_{\AdS_2}+(\d c^{(2)}-\chi\text{d}b^{(2)})\wedge \vol_{\text{S}^2}.
\end{split}	
\end{equation}

The existence of sensible regular solutions imposes the following conditions on the functions $A$, $B$, $H$ and $K$,
 \begin{itemize}
    \item The harmonic functions $A+\bar{A},B+\bar{B}$ and $K$ must have common singularities.
    \item No singular points should appear in the bulk of  the Riemann surface $\Sigma$.
    \item The functions $A+\bar{A},K$ and $H$ cannot have any zero in the bulk of the Riemann surface.
    \item The holomorphic functions $B$ and $\partial_zH$ must have common zeros.
 \end{itemize}
The previous conditions guarantee a non-vanishing  and finite everywhere $f_1$ (except at isolated singular points at the boundary), a finite $f_2$ in the interior of the Riemann surface and vanishing at the boundary, and, finally, finite and non-vanishing $f_3$ and $e^{2\Phi}$ functions everywhere on the Riemann surface, including the boundary.
  
 
The equations in \eqref{mmbbx} can be inverted to find $A$, $B$, $H$ and $K$ in terms of $f_i$ ($i=1,\dots, 3$), $\chi$ and $\Phi$. One finds two possibilities, that we will refer as the ``plus and minus solutions''\footnote{The ``plus solution'' corresponds to our AdS$_2$ backgrounds and the `` minus solutions'' to the AdS$_2$ geometries of  \cite{Lozano:2020txg}. Both these solutions are related through an analytical continuation, as explained around eq.(\ref{diagramaeq}).},
\begin{eqnarray}
	\label{kp-gen}
\text{Sol}_+:\; &&H = f_1f_2f_3^2,\quad K_{+}=\frac{f_1f_3^4}{2f_2},\quad A_{+}=\frac{f_1}{f_2}e^{2\Phi}-i\chi,\quad B_{+}=\frac{e^{\Phi}f_3^2}{2f_2}\sqrt{f_1^2-f_2^2}\;,\\
	\label{km-gen}
\text{Sol}_-:\; &&H = f_1f_2f_3^2, \quad K_{-}=\frac{f_2f_3^4}{2f_1},\quad A_{-}=\frac{f_2}{f_1}e^{2\Phi}-i\chi,\quad B_{-}=i\;\frac{e^{\Phi}f_3^2}{2f_1}\sqrt{f_1^2-f_2^2}.
\end{eqnarray}
Inserting the ``plus-solution'', equation \eqref{kp-gen},  or the ``minus-solution'', equation \eqref{km-gen}, in the first expression of \eqref{four-potential} one obtains, in both cases, the function associated to the 4-form potential,
\begin{eqnarray}
	C_{\text{CY}_2}=-2\tilde{K}, 
\end{eqnarray}
where $\tilde{K}$ is the harmonic function conjugate to $K$, according the footnote \ref{conjugatefoot}.

In the next subsection we obtain the harmonic and holomorphic functions that give rise to our backgrounds in eqs. \eqref{NS sector-B}-\eqref{RR sector lower rank fluxes}, as well as to the geometries in \cite{Lozano:2020txg}.

\subsection{Our AdS$_2$ geometries and the ``plus-solution''}\label{local-matching}

\paragraph{} 
In order to compare the generic backgrounds given by eqs.\eqref{NS sector-B}-\eqref{RR sector lower rank fluxes} with the solutions in \cite{Chiodaroli:2009yw, Chiodaroli:2009xh} we express our solutions in Einstein frame, to agree with their conventions. We obtain,
\begin{eqnarray}
\begin{split}
\label{R-MetricB}
f_1^2=\frac{u}{\sqrt{2}}\left(\frac{\hat{h}_4h_8^3}{(4\hat{h}_4h_8-(u')^2)^3}\right)^{1/4},\qquad f_2^2=\frac{u}{\sqrt{2^{5}}}\left(\frac{4\hat{h}_4h_8-(u')^2}{\hat{h}_4^3h_8}\right)^{1/4},\\
 f_3^2=\left(\frac{\hat{h}_4(4\hat{h}_4h_8-(u')^2)}{2^2h_8}\right)^{1/4},\qquad e^{2\Phi}=e^{-\phi}=\frac{1}{2}\sqrt{\frac{h_8}{\hat{h}_4}}	\sqrt{4\hat{h}_4h_8-(u')^2},\\
\chi=h_8'\psi, \qquad \tilde{\rho}^2=\frac{1}{\sqrt{2}u}
	\left(\hat{h}_4h_8^3(4\hat{h}_4h_8-(u')^2)\right)^{1/4},\qquad C_{\text{CY}_2}=-h_4'\psi.
\end{split}
\end{eqnarray}
We emphasise that these expressions are valid at the points where $h_8''=\hat{h}_4''=0$.

\begin{figure}[h]
\centering
\includegraphics[scale=0.55]{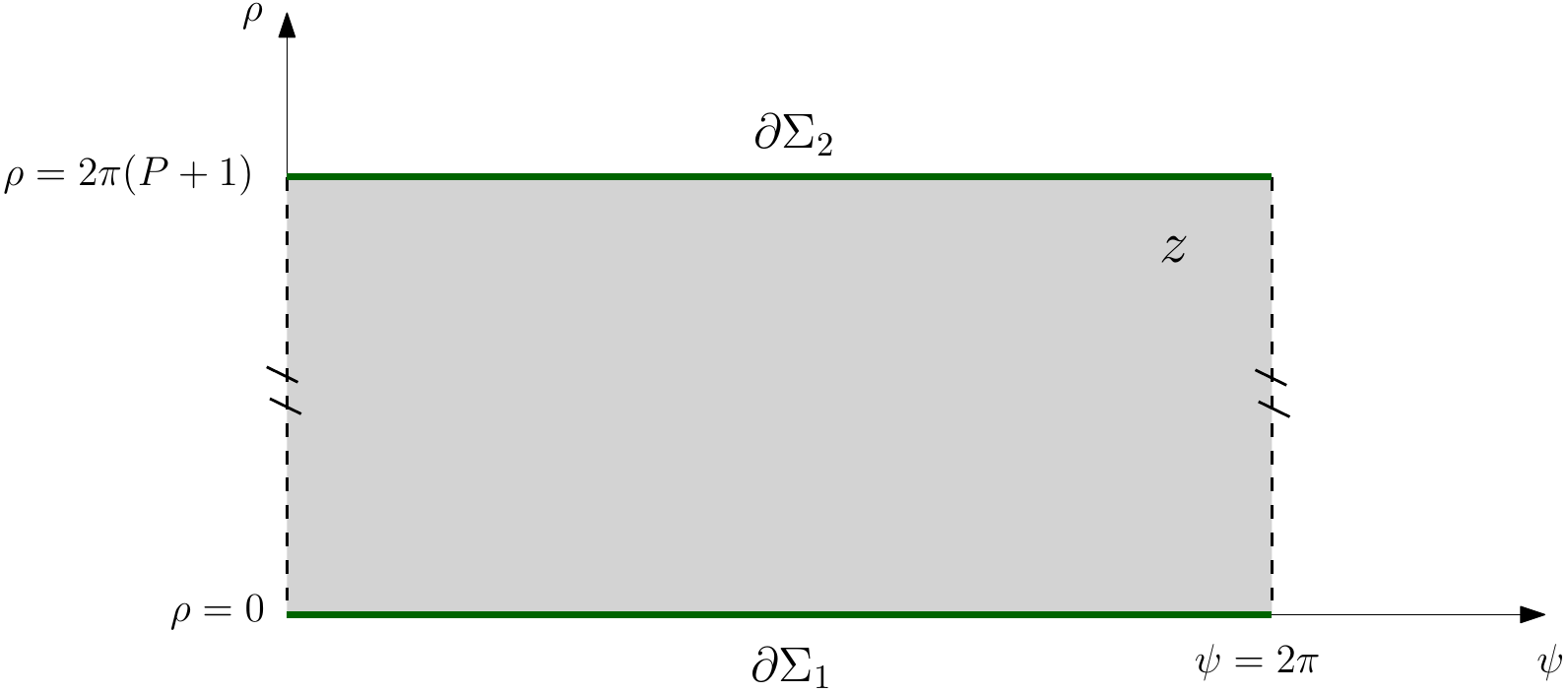}
\caption{Riemann surface associated to our AdS$_2$ geometries. Given the periodicity of $\psi$ it defines an annulus.}
\label{Strip-ABack}
\end{figure}

We take the $\rho$ and $\psi$ coordinates to define the real and imaginary parts of the $z$ variable. With this parametrisation, $\Sigma$ is an annulus, defined in the complex plane (see Figure \ref{Strip-ABack}),
\begin{equation}\label{StripA}
z=\psi + i \rho \qquad \text{with} \qquad \psi \in [0,2\pi]  \qquad \text{and} \qquad \rho \in [0,2\pi(P+1)] \, .
\end{equation}
Locally, our solutions are defined by the three functions $u, \hat{h}_4, h_8$, which must be linear in $\rho$. We take
\begin{equation}
\label{linearfunctions}
u=u_0+u_1\rho, \qquad h_8 = \mu+\nu\rho, \qquad  \hat{h}_4= \alpha+\beta\rho .
\end{equation}
Substituting \eqref{R-MetricB} in \eqref{kp-gen} and taking into account (\ref{linearfunctions}), we find 
for the functions $A, B, H, K$,     
\begin{equation}
\begin{split}
\label{A-BT}
	A=&h_8-i\psi h_8'=\mu-i\nu z , \qquad
	B=\frac{u'}{4}=\frac{u_1}{4} \, ,\\
	H=&\frac{u}{4}=\frac{u_0}{4}-i \frac{u_1}{8}(z-\bar{z}) \, ,\qquad
	K=\frac{\hat{h}_4}{2}=\frac{\alpha}{2}-i\frac{\beta}{4}(z-\bar{z}) \, .
	\end{split}
\end{equation}
It is easy to check that $H$, $K$, $A+\bar{A}$ and $B+\bar{B}$ are harmonic functions and $A$ and $B$ are holomorphic. 
The harmonic function $ \tilde{K}$ reads, in turn,
\begin{equation}
\label{k-ATHC}
	\tilde{K}=-\frac{\hat{h}_4'}{4}(z+\bar{z})= -\frac{\beta}{{4}}(z+\bar{z}) \,,
\end{equation}
which is the harmonic function conjugate to the expression for $K$ in \eqref{A-BT}.


From the equations \eqref{potential-b1}, and using \eqref{A-BT}, we can then obtain the harmonic functions and potentials associated with the NS-NS three-form,
\begin{equation}
\begin{split}
	&h_1=-\frac{i}{4}(z-\bar{z}) \, , \qquad \tilde{h}_1 = - \frac{1}{4}(z +\bar{z}) \, ,\\
	&b^{(1)} = \frac{u_1(2u_0-i u_1(z-\bar{z}))}{u_1^2+(2i\alpha+(z-\bar{z})\beta)(2i\mu+(z-\bar{z})\nu)} - h_1 \, , \qquad b^{(2)}=-\frac{1}{4}(z+\bar{z}) \, ,
	\end{split}
\end{equation}
as well as those associated with the RR three-form,
\begin{equation}
\begin{split}
&h_2=-\frac{\nu}{8}(z^2+\bar{z}^2)-\frac{i}{4}\mu(z-\bar{z}) \, ,  \qquad \tilde{h}_2 = i \frac{ \nu}{8}(z^2 - \bar{z}^2) -\frac{\mu}{4}(z+\bar{z})\,,\\
	&c^{(1)} = \frac{u_1(2u_0-i u_1(z-\bar{z}))(z+\bar{z})\nu}{8(u_1^2+(2i\alpha+(z-\bar{z})\beta)(2i\mu+(z-\bar{z})\nu))} + \tilde{h}_2 \, , \qquad c^{(2)} =-\frac{u_1 (2i u_0+u_1(z-\bar{z}))}{8 (\beta (z-\bar{z}) + 2 i\alpha)}+h_2  \, .
	\end{split}
\end{equation}
From these expressions we can  recover  $H_3$ and $F_3$ as given in eqs. \eqref{NS sector-B}-\eqref{RR sector lower rank fluxes}. Note that $h_i$ and $\tilde{h}_i$ are  harmonic functions conjugate to each other.

We have thus shown that our solutions can be obtained, locally, from the class of solutions constructed in  \cite{Chiodaroli:2009yw}. Note that in our analysis we have implicitly assumed that $h_8''=0$ and $\hat{h}_4''=0$ also hold globally. This is necessary in order to match the axion and the 4-form RR potential given in (\ref{R-MetricB}). This assumption --translated to our geometries-- indicates that we are not allowing for D7 and D3 brane sources, according to equations \eqref{caxa2}-\eqref{nana}. This agrees with the analysis in  \cite{Chiodaroli:2009yw}, which does not include either these types of branes. 

We will show in subsection \ref{annulus1} that D3-brane sources can be included in the two boundaries of the annulus following the formalism for the annulus derived in \cite{Chiodaroli:2009xh}. This  allows to recover the solutions in our class where D3-branes terminate the space at $\rho=2\pi (P+1)$. Quite surprisingly, we will also see that, even if not included in the analysis in \cite{Chiodaroli:2009xh}, D7-brane sources can also be allowed at the end of the space. We will show that they also manifest as (smeared) singularities of the basic harmonic function defined in the annulus in   \cite{Chiodaroli:2009xh}.

Before that, we show in the next subsection that the AdS$_2$ geometries found in \cite{Lozano:2020txg}, that we will refer as LNRS geometries, fit as well in the CGK class.



\subsection{The LNRS geometries and the ``minus-solution''}
\label{LNRS-section}

\paragraph{} As we already mentioned in section \ref{B-type}, our class of geometries can be obtained through a double analytic continuation from the AdS$_2$ solutions studied in \cite{Lozano:2020txg}. In this section we show that the latter fit within the class of solutions referred as ``minus solutions''  in \cite{Chiodaroli:2009yw}.       

The warping factors, dilaton, axion and RR 4-form potential associated to the AdS$_2$ geometries constructed in \cite{Lozano:2020txg} (in Einstein frame) are given by,
\begin{eqnarray}
\begin{split}
\label{WF-A}
f_1^2=\frac{u}{\sqrt{2^5}}\left(\frac{4\hat{h}_4h_8+(u')^2}{\hat{h}_4^3h_8}\right)^{1/4} \, , \qquad f_2^2=\frac{u}{\sqrt{2}}\left(\frac{\hat{h}_4h_8^3}{(4\hat{h}_4h_8+(u')^2)^3}\right)^{1/4} \, ,\\
f_3^2=\frac{1}{\sqrt{2}}\left(\frac{\hat{h}_4(4\hat{h}_4h_8+(u')^2)}{h_8}\right)^{1/4} \, , \qquad e^{2\Phi}=e^{-\phi}=\frac{1}{2}\sqrt{\frac{h_8}{\hat{h}_4}} \sqrt{4\hat{h}_4h_8+(u')^2} \, , \\
\chi=h_8'\psi \, , \qquad \tilde{\rho}^2=\frac{1}{\sqrt{2}u} \left(\hat{h}_4h_8^3(4\hat{h}_4h_8+(u')^2)\right)^{1/4} ,\qquad C_{\text{CY}_2}=-h_4'\psi\, .
\end{split}
\end{eqnarray}
The Riemann surface is the same one defined in equation \eqref{StripA} and Figure \ref{Strip-ABack}, and, as in the previous subsection, we are also taking  $h_8''=0$ and $\hat{h}_4''=0$ globally, i.e. solutions without D7 and D3 brane sources. This is needed to obtain the axion and RR 4-form potential of the previous equations.

In this case the matching with the solutions in \cite{Chiodaroli:2009yw} is with the 
``minus-solutions'' defined by equation \eqref{km-gen}. Taking into account \eqref{linearfunctions}, the harmonic and holomorphic functions read,
\begin{equation}
\begin{split}
\label{A-AT}
	A=h_8-i\psi h_8'=\mu-i\;\nu z  \, ,&\qquad
	B=i\;\frac{u'}{4}=i\;\frac{u_1}{4} \, ,\\
	H=\frac{u}{4}=\frac{u_0}{4}-i\frac{u_1}{8}(z-\bar{z})  \, ,&\qquad
	K=\frac{\hat{h}_4}{2}=\frac{\alpha}{2}-i\;\frac{\beta}{4}(z-\bar{z})   \, .
	\end{split}
\end{equation}
As in the previous subsection, the functions $H$, $K$, $A + \bar{A}$ $B + \bar{B}$ are harmonic and $A$ and $B$ holomorphic. The harmonic function $\tilde{K}$ reads exactly as in \eqref{k-ATHC}.

In turn, the harmonic functions that give rise to the NS-NS and RR three-forms read,
\begin{eqnarray}
\begin{split}
&h_1=-\frac{1}{4}(z+\bar{z}), \qquad \tilde{h}_1=\frac{i}{4}(z-\bar{z}),\\
	&h_2=-\frac{\mu}{4}(z+\bar{z})+i\frac{\nu}{8}(z^2-\bar{z}^2),\qquad \tilde{h}_2=i\frac{\mu}{4}(z-\bar{z})+\frac{\nu}{8}(z^2+\bar{z}^2),\\
&b^{(1)} = \frac{1}{4}(z + \bar{z}) \, ,\qquad 	b^{(2)}=\frac{u_1(2u_0-iu_1(z-\bar{z}))}{4(u_1^2-(2i\mu+\nu(z-\bar{z}))(2i\alpha+\beta(z-\bar{z})))} + \tilde{h}_1 \, ,\\
&c^{(1)} = -\frac{u_1 (u_1(z-\bar{z}) +2 i u_0)}{8(2i\alpha+\beta (z-\bar{z} ))} +\tilde{h}_2 \, , \qquad c^{(2)} = \frac{\nu u_1(2u_0-iu_1(z-\bar{z}))(z+\bar{z})}{8(u_1^2-(2i\mu+\nu(z-\bar{z}))(2i\alpha+\beta(z-\bar{z})))} + h_2 \, .
\end{split}
\end{eqnarray}
From these expressions we recover the NS-NS and RR field strengths, $H_3$ and $F_3$, of the solutions in \cite{Lozano:2020txg}.   
  

 \subsection{The annulus}\label{annulus1}
\paragraph{}
As we have already mentioned, the class of solutions constructed in \cite{Chiodaroli:2009yw}  have vanishing D3 and D7-brane charges. Those solutions have a Riemann surface with a single boundary component. 
In the follow-up paper \cite{Chiodaroli:2009xh}, the authors constructed solutions in which the Riemann surface $\Sigma$  has an arbitrary number of boundaries and non-vanishing D3 brane charges. The D3-branes occur as poles of a basic harmonic function at the boundaries.
In this section we consider the simplest case of a Riemann surface with two disconnected boundary components, namely the annulus.  We will then see in subsection \ref{annulus2} that we can recover the solutions with D3-brane sources at  $\rho=2\pi (P+1)$, the end of the space. Quite surprisingly, we will see that D7-branes seem also allowed at the end of the space. 

The annulus is defined as,
\begin{eqnarray}\label{annulus}
\Sigma\equiv\left\{w\in C,0\leq \text{Re}(w)	\leq 1,0\leq \text{Im}(w)	\leq\frac{t}{2}\right\},
\end{eqnarray}
with $t\in \mathbb{R}^+$. The points $w+1$ and $w$ are identified, thus giving the topology of an annulus. 
     Its two boundaries, $\partial\Sigma_{1,2}$, are located at Im$(w)=0$ and Im$(w)=\frac{t}{2}$. The annulus can be constructed from a double surface $\hat{\Sigma}$, which is defined as a rectangular torus with periods 1 and $\tau$, where $\tau$ is a purely imaginary parameter, $\tau=i t$. The original surface $\Sigma$ is obtained as the quotient  $\Sigma=\hat{\Sigma}/\mathcal{J}$ where $\mathcal{J}(z)=\bar{z}$.

The construction of the solutions for the annulus in \cite{Chiodaroli:2009xh} proceeds in three steps. First, a basic harmonic function with singularities and suitable boundary conditions is constructed. Second, the harmonic functions, $A+\bar{A}$, $H$ and $K$, are expressed as linear superpositions of the basic harmonic function, evaluated at the various poles in the two boundaries. Finally, the meromorphic function $B$ is constructed such that it satisfies certain regularity conditions. Some of these conditions come from imposing that the solutions asymptote locally to the AdS$_3\times \text{S}^3\times\text{CY}_2$ background. These regularity conditions will not be satisfied by our solutions, first because they do not asymptote to this geometry and, second, because the D3-branes (also the D7-branes) are smeared in the $\psi$ direction. This introduces significant changes in the regularity analysis. For this reason we will not give a detailed account of the regularity conditions imposed in \cite{Chiodaroli:2009xh}. We will see however in the next subsection that our solutions can still be recovered from the general formalism in \cite{Chiodaroli:2009xh} in an appropriate limit. 


The construction in \cite{Chiodaroli:2009xh} of the basic harmonic function is carried out in terms of elliptic functions and their related Jacobi theta function of the first kind,
    \begin{eqnarray}\label{theta1}
    \theta_{1}(w|\tau)=2\sum_{n=0}^{\infty}(-1)^n	e^{i\pi\tau(n+\frac{1}{2})^2}\sin[(2n+1)w],
    \end{eqnarray}
 as follows,
 \begin{eqnarray}\label{basic}
h_0(w,\bar{w})=i\left(\frac{\partial_{w}\theta_{1}(\pi w|\tau)}{\theta_{1}(\pi w|\tau)}+\frac{2\pi iw}{\tau}\right)+\text{c.c.}
     \end{eqnarray}
This function has the following simple properties: it has a single simple pole on $\partial \Sigma$, it satisfies Dirichlet conditions away from the pole, and it is positive in the interior of $\Sigma$. 
     
Notice that $h_0(w,\bar{w})$ has a singularity at $w=0$, on the first boundary. This pole can be shifted to any point on $\partial\Sigma_1$ by a real translation, so that $h_0(w-x,\bar{w}-x)$ has a singularity at $w=x$. Instead, to obtain the harmonic functions with singularities at $\partial\Sigma_2$ one needs to define,
     \begin{eqnarray}\label{wp}
     w'\equiv\frac{\tau}{2}	-w.
     \end{eqnarray}
Then the function $h_0(w'+y,\bar{w}'+y)$ has a pole at $w'=-y$, on the second boundary, for a real $y$. In other words the pole is localised at $w=y+it/2$.
     
 In the annulus the harmonic functions $A+\bar{A}$, $B+\bar{B}$, $H$ and $K$ are expressed as linear combinations of $h_0$ harmonic functions with poles on both boundaries,
 \begin{equation}
 \begin{split}
 A+\bar{A}=&\sum_{\ell_A=1}	^{M_A}r_{\ell_A}h_0(w-x_{\ell_A},\bar{w}-x_{\ell_A})+\sum_{j_A=1}	^{M'_A}r'_{j_A}h_0(w'+y_{j_A},\bar{w}'+y_{j_A}),
 \\
 B+\bar{B}=&\sum_{\ell_B=1}	^{M_B}r_{\ell_B}h_0(w-x_{\ell_B},\bar{w}-x_{\ell_B})+\sum_{j_B=1}	^{M'_B}r'_{j_B}h_0(w'+y_{j_B},\bar{w}'+y_{j_B}),
 \\
 H=&\sum_{\ell_H=1}	^{M_H}r_{\ell_H}h_0(w-x_{\ell_H},\bar{w}-x_{\ell_H})+\sum_{j_H=1}	^{M'_H}r'_{j_H}h_0(w'+y_{j_H},\bar{w}'+y_{j_H}),
 \\
  K=&\sum_{\ell_K=1}	^{M_K}r_{\ell_K}h_0(w-x_{\ell_K},\bar{w}-x_{\ell_K})+\sum_{j_K=1}	^{M'_K}r'_{j_K}h_0(w'+y_{j_K},\bar{w}'+y_{j_K})\label{kann}.
  \end{split}
 \end{equation}
 Each harmonic function is taken to have $M_i$ poles $x_{\ell_i}$ with $\ell_{i}=1,...,M_i$ on $\partial\Sigma_1$, and $M_i'$ poles $y_{j_i}$ with $j_{i}=1,...,M'_i$ on $\partial\Sigma_2$. The corresponding residues are $r_{\ell_i}$ and $r_{j_i}$.  
 
 In addition to the regularity conditions given in subsection \ref{Gutperlething}, the harmonic functions \eqref{kann} satisfy  an extra condition coming from the requirement  that $e^{4\Phi}>0$. Namely, $(A+\bar{A})K-(B+\bar{B})^2>0$ must be obeyed throughout $\Sigma$. Furthermore, in this language the first regularity condition can be written in terms of the residues as $r_Ar_K=r_B^2$.

\subsection{Zoom-in to our solutions}\label{annulus2}

\paragraph{} In this subsection we show that it is possible to recover well-defined global solutions with source branes at the ends of the space from the general analysis above for the annulus. These solutions do not satisfy most of the regularity conditions imposed in  \cite{Chiodaroli:2009yw,Chiodaroli:2009xh}, and, moreover,  contain not only D3 but also D7-brane sources at the ends of the space. Still, we will be able to recover them in a particular limit from the formalism in  \cite{Chiodaroli:2009xh}. 

As we have already stressed, the choice of constants in the general $\hat{h}_4$ and $h_8$ functions defined by equations (\ref{profileh4sp}) and (\ref{profileh8sp}) allows for discontinuities in the RR sector of our backgrounds at each $\rho=2\pi k$ value, with $k=1,\dots, P$. The discontinuities in $\hat{h}_4'$ are interpreted as generated by D3-brane sources,  while the discontinuities in $h_8'$ are interpreted as generated by D7-branes. Both types of branes are smeared in the $\psi$ direction. The space is terminated in the $\rho$ direction by imposing that $\hat{h}_4= h_8=0$ at $\rho=0, 2\pi (P+1)$. 
When $u=$ constant the closure of the space by setting $\hat{h}_4=h_8=0$ generates D3 and D7 sources, in the boundary of the space, as explained around eq.(\ref{O3O7brane-singularity}).

Instead, in the general discussion for the annulus in \cite{Chiodaroli:2009xh} the D3-branes occur as poles of a basic 
harmonic function at its two boundaries. The basic harmonic function must however be regular in the interior. Therefore, in order to fit in the discussion for the annulus we need continuous $\hat{h}_4'$ and $h_8'$ functions. This is imposed taking
\begin{equation}
\label{increasing2}
\beta_k\equiv \beta, \qquad \nu_k\equiv \nu, \qquad \text{for} \qquad k=0,1,\dots, P,
\end{equation}
in (\ref{profileh4sp}), (\ref{profileh8sp}), which implies
\begin{equation}
\label{increasing}
\alpha_k=k\, \beta, \qquad \mu_k=k\, \nu, \qquad \text{for} \qquad k=0,\dots, P.
\end{equation}
The solutions are then defined by the functions
\begin{equation}
\hat{h}_4=\frac{\beta}{2\pi}\, \rho, \qquad h_8=\frac{\nu}{2\pi}\,\rho
\end{equation}
at all $\rho$-intervals. Yet, the closure of the space at $\rho=2\pi (P+1)$ requires that $(P+1)\beta$ D3-branes and $(P+1)\nu$ D7-branes are present at the end of the space. Instead of closing the space by introducing sources as we did with the choice of $\hat{h}_4$ and $h_8$ functions given by (\ref{profileh4sp}) and (\ref{profileh8sp}), these branes will be automatically present at the end of the space in the annulus construction. 



Let us now see how these solutions arise from the general formalism in \cite{Chiodaroli:2009xh}. We take the annulus in \eqref{annulus} as defined from,
\begin{eqnarray}\label{w}
\begin{split}
&w=\frac{z}{2\pi}=\tilde{\psi}+i\tilde{\rho},	\quad &\text{with}\qquad\tilde{\psi}=\frac{\psi}{2\pi}, \quad \tilde{\rho}=\frac{\rho}{2\pi}.
\end{split}
\end{eqnarray}
Then $\tilde\psi\in [0,1]$ and the parameter $t$ in the definition of the annulus is $t=2(P+1)$. 
As recalled in section \ref{B-type}, our class of solutions is valid when $P$ is large. This allows us to approximate the Jacobi theta function introduced in (\ref{theta1}) by its 
asymptotic expansion when $t\to\infty$,
 \begin{eqnarray}\label{theta-ex1}
    \theta_{1}(\pi w|\tau)|_{t\to \infty}\approx 2e^{-\frac{\pi}{4}t}\sin{\pi w}\approx i e^{-\frac{\pi}{4}t}e^{-i\pi w}.
    \end{eqnarray}
    This approximation will be key in showing the matching with our solutions. Indeed, 
in this approximation it is easy to see that the basic harmonic function defined by \eqref{basic} reads,
\begin{eqnarray}\label{h_0-exp}
 h_0(w,\bar{w})\approx	2\pi+\frac{i\pi}{P+1}(w-\bar{w}).
 \end{eqnarray}
This gives, at the two boundaries $\partial\Sigma_1$ and $\partial\Sigma_2$, 
   \begin{eqnarray}\begin{split}\label{h_0boundaries}
& h_0(w-x_{\ell_i},\bar{w}-x_{\ell_i})\approx	2\pi+\frac{i\pi}{P+1}(w-\bar{w}),\\
& h_0(w'+y_{j_i},\bar{w}'+y_{j_i})\approx-\frac{i\pi}{P+1}(w-\bar{w}),
\end{split}
 \end{eqnarray}
 respectively, where for the second boundary we have used the relation \eqref{wp}. These expressions are thus independent of the positions of the poles at both boundaries. This is in agreement with the fact that our D3/D7 branes are smeared in the $\psi$-direction. We then
get for the harmonic functions in eq.\eqref{kann},
 \begin{equation}
 \begin{split}
 A+\bar{A}=&2\pi\sum_{\ell_A=1}	^{M_A}r_{\ell_A}-\frac{i\pi}{P+1}(w-\bar{w})\left(\sum_{j_A=1}^{M'_A}r'_{j_A}-\sum_{\ell_A=1}	^{M_A}r_{\ell_A}\right),
 \\
 B+\bar{B}=&2\pi\sum_{\ell_B=1}	^{M_B}r_{\ell_B}-\frac{i\pi}{P+1}(w-\bar{w})\left(\sum_{j_B=1}	^{M'_B}r'_{j_B}-\sum_{\ell_B=1}	^{M_B}r_{\ell_B}\right),
 \\
 H=&2\pi\sum_{\ell_H=1}	^{M_H}r_{\ell_H}-\frac{i\pi}{P+1}(w-\bar{w})\left(\sum_{j_H=1}	^{M'_H}r'_{j_H}-\sum_{\ell_H=1}	^{M_H}r_{\ell_H}\right),
 \\
  K=&2\pi\sum_{\ell_K=1}	^{M_K}r_{\ell_K}-\frac{i\pi}{P+1}(w-\bar{w})\left(\sum_{j_K=1}	^{M'_K}r'_{j_K}-\sum_{\ell_K=1}	^{M_K}r_{\ell_K}\right)\label{kann2}.
  \end{split}
  \end{equation}
 
In order to match these expressions with the expressions for $A+\bar{A}$ and $K$ given in eq.(\ref{A-BT}) we take into account that $w=z/(2\pi)$, and we obtain,
 \begin{equation}\label{Aannulus}
  \sum_{\ell_A=1}	^{M_A}r_{\ell_A}=0 \qquad \text{and} \qquad \sum_{j_A=1}^{M'_A}r'_{j_A}=\frac{(P+1)\nu}{4\pi},
\end{equation} 
for the matching of $A+\bar{A}$, and
 \begin{equation}\label{kannulus}
  \sum_{\ell_K=1}	^{M_K}r_{\ell_K}=0 \qquad \text{and} \qquad \sum_{j_K=1}^{M'_K}r'_{j_K}=\frac{(P+1)\beta}{\pi},
\end{equation} 
for the matching of $K$. Rescaling the residues as\footnote{Note that a rescaling is also necessary in order to interpret the residues of the solutions in \cite{DHoker:2017mds} as charges of $(p,q)$ 5-branes.}  
\begin{equation}
	r'_{j_A}\to 2r'_{j_A},\qquad r'_{j_K}\to \frac{r'_{j_K}}{4}.
\end{equation}
and replacing the sums by
\begin{equation}
 \sum_{j_A=1}^{M'_A}r'_{j_A}\rightarrow \frac{1}{2\pi}\int_0^{2\pi}  dr'_{j_A},
 \end{equation}
as implied by the smearing of the branes in the $\psi$-direction, we can finally interpret the residues as the charge-densities of D7 and D3 brane sources at both boundaries of the annulus. We would like to stress that even if the general formalism in \cite{Chiodaroli:2009xh} does not account for D7-branes at the boundaries of the annulus, we have associated these to the (smeared) poles of the basic harmonic function for $A+\bar{A}$. The analysis goes in complete parallelism to the analysis of the residues and poles of the $K$ function, associated to the D3-brane sources at both boundaries of the annulus.  It is unclear to us the precise reason why this seems to work in the presence of D7-branes. 


 Finally, from the matching of the $B+\bar{B}$ and $H$ functions we find
 \begin{equation}
 \sum_{\ell_B=1}	^{M_B}r_{\ell_B}=\sum_{j_B=1}	^{M'_B}r'_{j_B}=0 \qquad \text{and} \qquad \sum_{\ell_H=1}	^{M_H}r_{\ell_H}=\sum_{j_H=1}	^{M'_H}r'_{j_H}=\frac{u_0}{8\pi}.
 \end{equation}
 These expressions do not seem to have however a direct interpretation in terms of charges of our solutions. 
  
  The previous analysis holds true as well for the LNRS backgrounds discussed in  \cite{Lozano:2020txg}. The matching of the $A+\bar{A}$, $H$ and $K$ functions is valid for both solutions, while the harmonic function $B+\bar{B}$ vanishes. Again, there are smeared D3 and D7-branes at the end of the space with the same relations between residues and charges.

 \section{A new class of AdS$_2\times \text{S}^2\times \text{CY}_2\times \Sigma$ solutions with $\Sigma$ an infinite strip}\label{NATD}
 
\paragraph{}
In this section we construct a new class of AdS$_2$ solutions to Type IIB supergravity with 8 supercharges by acting with non-Abelian T-duality (NATD) on the AdS$_3\times$S$^2\times$CY$_2\times$I$_\rho$ solutions obtained in \cite{Lozano:2019emq}. The non-Abelian T-duality transformation is performed with respect to a freely acting SL$(2,\mathbb{R})$ isometry group of the AdS$_3$ subspace. This transformation gives rise to a new class of solutions in which the AdS$_3$ subspace is replaced by AdS$_2$ times an interval. These solutions fit in the classification of \cite{Chiodaroli:2009yw} for a Riemann surface with a single boundary, equivalent to an infinite strip. 

\subsection{NATD of the AdS$_3\times$S$^2\times$CY$_2$ solutions}

\paragraph{}The study of NATD as a solution generating technique of supergravity was initiated in    \cite{Sfetsos:2010uq}. Further works include \cite{Lozano:2011kb, Itsios:2012dc, Itsios:2012zv, Itsios:2013wd}. In all these examples the dualisation took place with respect to a freely acting SU(2) subgroup of the entire symmetry group of the solutions. Instead, in this section we perform the non-Abelian T-duality transformation with respect to one of the freely acting SL$(2,\mathbb{R})$ isometry groups of the AdS$_3$ subspace. 

In order to perform the dualisation with respect to the SL$(2,\mathbb{R})$ isometry group we follow the derivation in \cite{Alvarez:1993qi}. We take the $sl(2,\mathbb{R})$ generators analytically continuing the $su(2)$ generators, as $t^a=\tau_a/\sqrt{2}$, with
\begin{eqnarray}
\tau_1=\left(\begin{matrix}
  &0 & i\\
  &i& 0
\end{matrix}\;\;\right),\qquad \tau_2=\left(\begin{matrix}
  &0 & -i\\
  &i& 0
\end{matrix}\;\;\right),	\qquad \tau_3=\left(\begin{matrix}
  &i & 0\\
  &0& -i
\end{matrix}\;\;\right).		
\end{eqnarray}
These generators satisfy,
\begin{equation}
\Tr(t^at^b)=(-1)^a\delta^{ab},\quad\quad\quad	[t^1,t^2]=i\sqrt{2}t^3,\quad	[t^2,t^3]=i\sqrt{2}t^1,\quad	[t^3,t^1]=-i\sqrt{2}t^2.
\end{equation}
A group element in the Euler parametrisation is given by,
\begin{equation}
g=e^{\frac{i}{2}\phi\tau_3}e^{\frac{i}{2}\theta\tau_2}e^{\frac{i}{2}\psi\tau_3},	
\qquad\text{where} \qquad 0\leq\theta\leq\pi,\;\; 0\leq\psi< \infty,\;\; 0\leq\phi< \infty,
\end{equation}
from which we write the left invariant one forms, $L^a=-i\Tr(t^ag^{-1}\text{d}g)$, in the following fashion,
\begin{eqnarray}
\begin{split}
\label{Maurer}
L^1=&\sinh{\psi}\text{d}\theta-\cosh{\psi}\sin{\theta}\text{d}\phi\\
L^2=&\cosh{\psi}\text{d}\theta-\sinh{\psi}\sin{\theta}\text{d}\phi\\
L^3=&-\cos{\theta}\text{d}\phi-\text{d}\psi.
\end{split}
\end{eqnarray}

The backgrounds in \cite{Lozano:2019emq} support an SL$(2,\mathbb{R})$ isometry such that the metric, the Kalb-Ramond field and the dilaton  can be written as\footnote{We have taken $g^{\mu\nu}=-\Tr(t^\mu t^\nu)$ to have signature $(+,-,+)$.},
\begin{equation}
\begin{split}
ds^2&=\frac{1}{4}g_{\mu\nu}(x)L^\mu L^\nu+G_{ij}(x)dx^idx^j,\qquad
B_2=
B_{ij}(x)dx^i\wedge dx^j, \qquad \phi=\phi(x),
\end{split}
\end{equation}
where $x^i$ are the coordinates in the internal manifold, for $i,j=1,2,...,7$, and $L^{\mu}$ are the forms given by \eqref{Maurer}. All the coordinate dependence on the SL$(2,\mathbb{R})$ group is contained in these forms. The subsequent details on how to technically compute the NATD transformation have been developed extensively in the literature \cite{Sfetsos:2010uq, Itsios:2013wd} (see these reference for more details). 

The geometries obtained through NATD with respect to a freely acting SL$(2,\mathbb{R})$ group on the AdS$_3$ of the solutions in \cite{Lozano:2019emq} are given by,
\begin{equation}
\begin{split}
\text{d}{s}_{st}^2&=\frac{u\sqrt{\hat{h}_4 h_8}}{4 r^2\hat{h}_4 h_8-u^2} r^2\text{d}s_{\text{AdS}_2}^2+\sqrt{\frac{\hat{h}_4}{h_8}}\text{d}s^2_{\text{CY}_2}+\frac{u\sqrt{\hat{h}_4 h_8} }{4 \hat{h}_4 h_8+u'^2}\text{d}s^2_{\text{S}^2}+\frac{\sqrt{\hat{h}_4 h_8} }{u}(\text{d}\rho ^2+\text{dr}^2),\\
e^{-2\phi}&=\frac{\left(4 r^2\hat{h}_4 h_8-u^2\right) \left(4 \hat{h}_4 h_8+u'^2\right)}{2^8\hat{h}_4^2},\\
B_2&=-\frac{2 r^3\hat{h}_4 h_8}{4 r^2\hat{h}_4 h_8-u^2}\text{vol}_{\text{AdS}_2}-\frac{4 \rho \hat{h}_4 h_8-u'(u-\rho  u')}{2 \left(4 \hat{h}_4 h_8+u'^2\right)} \text{vol}_{\text{S}^2}\label{B2NATD}.
\end{split}
\end{equation}
Additionally, the background is supported by the RR fluxes,
\begin{equation}
\begin{split}
F_1=&-\frac{r h'_8}{4}\text{d}r+\frac{1}{2^4}\left(4h_8+\partial_\rho \left[\frac{u u'}{\hat{h}_4}\right]\right)\text{d}\rho,\\
F_3=&\frac{h_8}{8 \left(4 \hat{h}_4 h_8+u'^2\right)}\left(\frac{\hat{h}_4'u^2}{\hat{h}_4}\text{d}\rho+rh_8\left(4\hat{h}_4+\partial_\rho \left[\frac{u u'}{h_8}\right]\right)\text{d}r\right)\wedge \text{vol}_{\text{S}^2}\\
&+\frac{r^2h_8}{8 \left(4 r^2\hat{h}_4 h_8-u^2\right)}\left(\frac{h_8'u^2}{h_8}\text{d}r-r\hat{h}_4\left(4h_8+\partial_\rho \left[\frac{u u'}{\hat{h}_4}\right]\right)\text{d}\rho\right)\wedge \text{vol}_{\text{AdS}_2},\label{F3NATD}\\
F_5=&\frac{1}{2^4}\left(4r\hat{h}_4'\text{d}r-\left(4\hat{h}_4+\partial_\rho \left[\frac{u u'}{h_8}\right]\right)\text{d}\rho\right)\wedge \text{vol}_{\text{CY}_2}\\
&-\frac{r^2u^2h_8^2}{2^4\left(4 r^2\hat{h}_4 h_8-u^2\right) \left(4 \hat{h}_4 h_8+u'^2\right)}\left(4 r\hat{h}_4'\text{d$\rho $}+\left(4\hat{h}_4+\partial_\rho \left[\frac{u u'}{h_8}\right]\right)\text{d$r$}\right)\wedge\text{vol}_{\text{AdS}_2}\wedge \text{vol}_{\text{S}^2},\\
F_7=&-\frac{\hat{h}_4}{8 \left(4\hat{h}_4 h_8+u'^2\right)}\left(\frac{h_8'u^2}{h_8}\text{d}\rho+r\hat{h}_4\left(4h_8+\partial_\rho \left[\frac{u u'}{\hat{h}_4}\right]\right)\text{d}r\right)\wedge \text{vol}_{\text{S}^2}\wedge \text{vol}_{\text{CY}_2}\\
&-\frac{r^2\hat{h}_4}{8 \left(4 r^2\hat{h}_4 h_8-u^2\right)}\left(\frac{\hat{h}_4'u^2}{\hat{h}_4}\text{d}r-rh_8\left(4\hat{h}_4+\partial_\rho \left[\frac{u u'}{h_8}\right]\right)\text{d}\rho\right)\wedge \text{vol}_{\text{AdS}_2}\wedge \text{vol}_{\text{CY}_2},\\
F_9=&\frac{r^2u^2\hat{h}_4^2}{4\left(4 r^2\hat{h}_4 h_8-u^2\right)\left(4 \hat{h}_4 h_8+u'^2\right)}\left(rh_8'\text{d}\rho+\frac{1}{4}\left(4h_8+ \partial_\rho \left[\frac{u u'}{\hat{h}_4}\right]\right)\text{d}r\right)\wedge \text{vol}_{\text{AdS}_2}\wedge \text{vol}_{\text{CY}_2}\wedge \text{vol}_{\text{S}^2}.
\end{split}
\end{equation}
The previous background is  a solution to the Type IIB supergravity EOM whenever $4 r^2\hat{h}_4 h_8-u^2>0$. Namely we get a well-defined geometry for 
\begin{equation}
r>r_0=\frac{u}{2\sqrt{\hat{h}_4 h_8}}.\label{r0natd}
\end{equation}

In the next section we show that a subset of the solutions defined by  \eqref{B2NATD} and \eqref{F3NATD} fit in the general classification of AdS$_2\times$S$^2\times$CY$_2\times\Sigma$ geometries given in \cite{Chiodaroli:2009yw} with $\Sigma$ an infinite strip.

\subsection{The NATD solution as a CGK geometry}

\paragraph{} In this section we discuss how the solutions given by \eqref{B2NATD}-\eqref{F3NATD}  fit in the class of CGK.     Going to Einstein frame we get the warp factors of the metric, dilaton and  axion, 
\begin{equation}
\begin{split}	
	&f_1^2=\frac{ur^2\sqrt{h_8}(4 \hat{h}_4 h_8+u'^2)^{1/4}}{4(4 \hat{h}_4 h_8r^2-u^2)^{3/4}},\quad
	f_2^2=\frac{u\sqrt{h_8}(4 \hat{h}_4 h_8r^2-u^2)^{1/4}}{4(4 \hat{h}_4 h_8+u'^2)^{3/4}},\\
	&f_3^2=\frac{(4 \hat{h}_4 h_8r^2-u^2)^{1/4}(4 \hat{h}_4 h_8+u'^2)^{1/4}}{4\sqrt{h_8}},\quad
	e^{2\Phi}=e^{-\phi}=\frac{\sqrt{(4 \hat{h}_4 h_8r^2-u^2)(4 \hat{h}_4 h_8+u'^2)}}{2^4\hat{h}_4},\\
	&\chi =\frac{1}{2^4}\left(2\nu(\rho^2-r^2)+4\mu\rho+\frac{u u'}{\hat{h}_4}\right),\quad \tilde{\rho}^2=\frac{\sqrt{h_8}(4 \hat{h}_4 h_8r^2-u^2)^{1/4}(4 \hat{h}_4 h_8+u'^2)^{1/4}}{2^2u}
	.
	\end{split}
\end{equation}
In $\chi$, the axion field, we have taken $h_8=\mu+\nu\rho$, with $\mu$, $\nu$ constants. 
This choice corresponds to backgrounds without D7-branes,
as those constructed in \cite{Chiodaroli:2009yw}. 

\begin{figure}[t]
\centering
\includegraphics[scale=0.6]{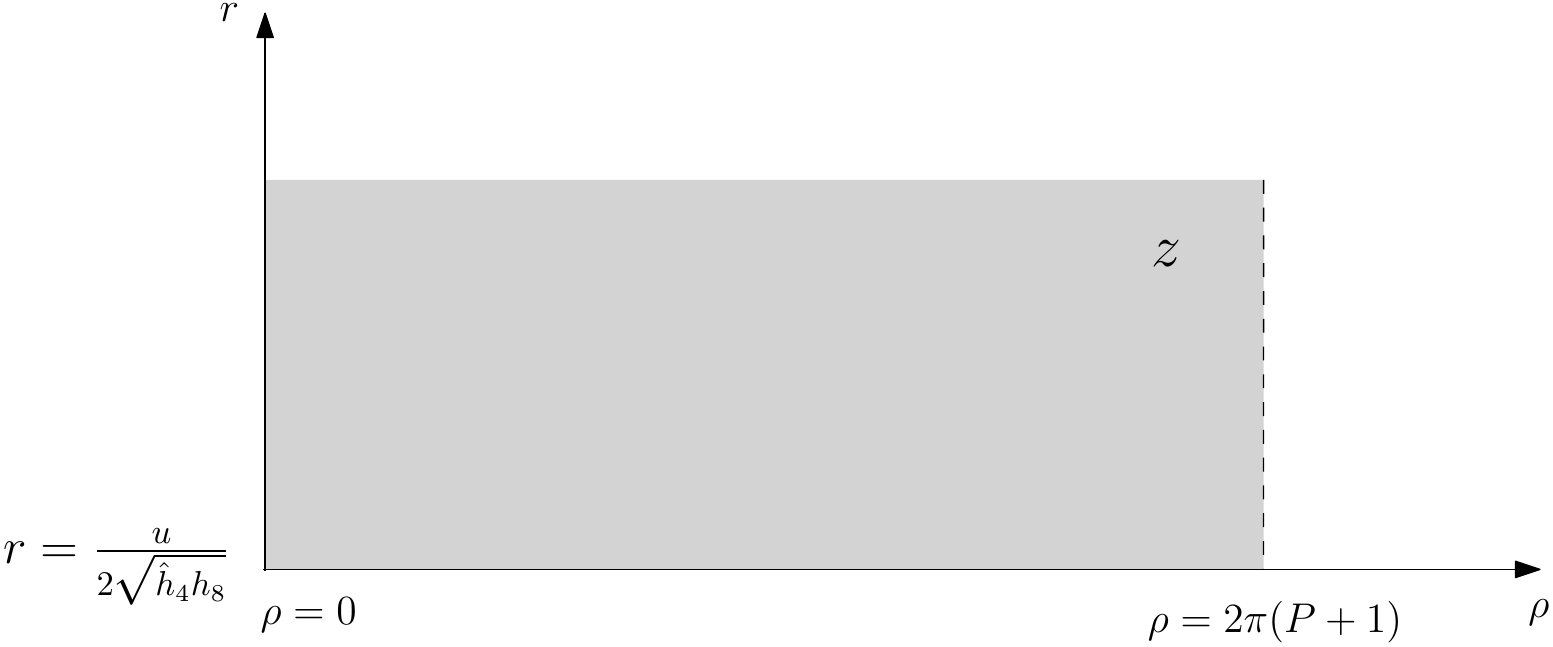}
\caption{Infinite strip associated to the NATD solution.}
\label{strip2}
\end{figure}

The 2d Riemann surface associated to the solutions is the strip depicted in Figure \ref{strip2}, parametrised as,
\begin{eqnarray}
	\label{StripD}
	z=\rho+i\;r\qquad
	\text{where}\qquad \rho\in[0,2\pi(P+1)]\qquad\text{and}\qquad r\in[r_0,\infty],
\end{eqnarray}
%
where the value of $r_0$ is determined in eq.(\ref{r0natd}).

Taking the 'plus-solution' defined by equation \eqref{kp-gen} we obtain the $A$, $B$, $H$ and $K$ functions in terms of the defining functions of our backgrounds, $\hat{h}_4$, $h_8$ and $u$, 
\begin{equation}
\begin{split}
A=\frac{1}{2^4}\left(4\mu(r-i\rho)+2i\nu(r-i\rho)^2+\frac{u'}{\hat{h}_4}(ru'-iu)\right),	\quad
B=\frac{1}{2^5}&\frac{\sqrt{4 \hat{h}_4 h_8+u'^2}\sqrt{u^2+r^2u'^2}}{\sqrt{\hat{h}_4 h_8}},\\
H=\frac{ru}{2^4},\qquad
K=\frac{r(4 \hat{h}_4 h_8+u'^2)}{2^5h_8}.
\end{split}
\end{equation}
We anticipate these functions are neither harmonic nor holomorphic. In order to ensure harmonicity -in $H$ and $K$- and holomorphicity -in $A$ and $B$- we need to choose $u'=0$. In that case we obtain,   
\begin{equation}
\begin{split}
	&A=\frac{4\mu(r-i\rho)+2i\nu(r-i\rho)^2}{2^4}=-iz\frac{2\mu+z\nu}{8},	\qquad
	B=\frac{u}{2^4}=\frac{u_0}{2^4},\\
	&H=\frac{ru}{2^4}=-i\frac{u_0}{2^5}(z-\bar{z}),\qquad
	K=\frac{r \hat{h}_4}{2^3}=-i\frac{(z-\bar{z})}{2^5}(\beta(z+\bar{z})+2\alpha),
	\end{split}
\end{equation}
where we have used \eqref{linearfunctions}. The harmonic function conjugated to $K$ is,
\begin{eqnarray}
	\tilde{K}=-\frac{1}{2^5}(\beta(z^2+\bar{z}^2)+2\alpha(z+\bar{z})),\qquad\text{with}\qquad C_{\text{CY}_2}=\frac{1}{2^3}(\beta(r^2-\rho^2)+2\alpha\rho).
\end{eqnarray}
Note that we have taken $\hat{h}_8=\alpha+\beta\rho$, with $\alpha$, $\beta$ constants, which corresponds to backgrounds without D3-branes, as those constructed in  \cite{Chiodaroli:2009yw}.

The functions associated to the complex three-form are,
\begin{equation}
\begin{split}
	&h_1=-\frac{i}{4}(z-\bar{z}),\qquad \tilde{h}_1=-\frac{1}{4}(z+\bar{z}),\\
	&h_2=-\frac{1}{2^5}\left(\mu(z^2+\bar{z}^2)+\frac{\nu}{3}(z^3+\bar{z}^3)\right),\quad \tilde{h}_2=\frac{i}{2^5}\left(\mu(z^2-\bar{z}^2)+\frac{\nu}{3}(z^3-\bar{z}^3)\right).
	\end{split}
\end{equation}
Notice that $h_i$ and $\tilde{h}_i$ are harmonic functions conjugate to each other. The potentials given in \eqref{potential-b1} are,
\begin{equation}
\begin{split}
&	b^{(1)}=-i\frac{u_0^2(z-\bar{z})}{(z-\bar{z})^2(\beta(z+\bar{z})+2\alpha)(\nu(z+\bar{z})+2\mu)+4u_0^2}-h_1,\qquad b^{(2)}=-\frac{1}{4}(z+\bar{z}),\\
&	c^{(1)}=-i\frac{u_0^2(z-\bar{z})(2\mu(z+\bar{z})+\nu(z^2+\bar{z}^2))}{16((z-\bar{z})^2(\beta(z+\bar{z})+2\alpha)(\nu(z+\bar{z})+2\mu)+4u_0^2)}+\tilde{h}_2,\\
& c^{(2)}=-\frac{u_0^2}{16(\beta(z+\bar{z})+2\alpha)}+h_2,
\end{split}
\end{equation}
 which agree with the expressions \eqref{B2NATD} and \eqref{F3NATD} for $u'=0$.

The previous analysis shows that the new class of solutions constructed through non-Abelian T-duality provide an explicit example of CGK geometries where the Riemann surface is an infinite strip. We will provide a more detailed global study of these solutions in a future publication.

  \section{Electric-magnetic charges and a minimisation principle}\label{maxwellcentralhol}
\paragraph{}
In this section we extend  two results discussed in \cite{Lozano:2020txg, Lozano:2020sae} to our new  infinite family of  AdS$_2$ solutions.  
\\
The first result  is a relation between the holographic central charge in eq.(\ref{hccharge-B}) and  an integral of the product of the electric and magnetic fluxes of the Dp-branes present in the background.  This relates the holographic central charge in Section \ref{hccharge}, computed purely in terms of the NS-NS sector of the background, with a calculation purely in terms of the Ramond-Ramond sector. 
\\
Furthermore, in section \ref{functional}, we explore this relation from a geometrical point of view. We define a quantity in terms of geometric forms  in our geometries and through an extremisation principle relate it to the holographic central charge in eq. (\ref{hccharge-B}). In summary, in this section we present a connection between  the holographic central charge,  the product of the electric and magnetic charges and an extremised functional.

\subsection{A relation between the holographic central charge and the RR fluxes}

\paragraph{} We provide a relation between the holographic central charge found in eq.(\ref{hccharge-B}) and the  fluxes of the Ramond-Ramond sector in eq \eqref{RR sector lower rank fluxes}.
Consider a Dp brane and the associated electric  $\hat{F}_{p+2}$ and magnetic $\hat{F}_{8-p}$ Page field strengths. We define the ``density of electric and magnetic charges'',  $\rho^e_{\text{Dp}}$ and $\rho^m_{\text{Dp}}$, as follows,
\begin{eqnarray}
& & \rho_{\text{Dp}}^{e}=\frac{1}{(2\pi)^p} \hat{F}_{p+2} ,\;\;\;\;\; \rho_{\text{Dp}}^{m}=\frac{1}{(2\pi)^{7-p}} \hat{F}_{8-p}.\label{mc}
\end{eqnarray}
From these we construct the quantity,
\begin{equation}	
\begin{split}
\label{sumdensities}
& \int \sum_{p=1,3,5,7} \rho_{\text{D}p}^e \rho_{\text{D}p}^m=\,\\
& = \frac{1}{\pi}\text{Vol}_{\text{AdS}_2}\left(\frac{\text{Vol}_{\text{CY$_2$}}} {16 \pi^4}\right) \int 	\text{d} \rho\left[\frac{4\hat{h}_4h_8-(u')^2}{8} + \frac{1}{16}\partial_\rho\left(u^2\frac{(h_4 h_8)'}{h_4h_8} \right)-\frac{u^2}{16}\left(\frac{\hat{h}_4''}{\hat{h}_4}+\frac{h_8''}{h_8}\right) \right] 
.
\end{split}
\end{equation}
In the absence of sources $\hat{h}_4''= h_8''=0$ and, up to a boundary term, this is proportional to the expression for the holographic central charge in equation \eqref{hccharge-B}. We explore below the contribution of the sources to this expression. Notice that eq.(\ref{sumdensities}) links the holographic central charge in eq.\eqref{hccharge-B}---a calculation purely in terms of the NS-NS sector---with one purely in terms of the Ramond-Ramond sector. 

\subsection{An action functional for the central charge}\label{functional}
 \paragraph{} Following the ideas of \cite{Couzens:2018wnk, Gauntlett:2018dpc} and the lead of the works \cite{Lozano:2020txg, Lozano:2020sae},  we construct a functional in terms of an integral of  forms defined in the internal space. Once such functional is extremised the holographic central charge in eq.(\ref{hccharge-B})  is recovered, up to a boundary term.  

 We define forms $J_i$ and ${\mathcal F}_i$ (for $i=1,3,5,7$) on the internal space $X_8=$[S$^2$, CY$_2$, S$_\psi$, I$_\rho$]. These forms are  inherited from the Page fluxes \eqref{fluxes23}\footnote{The same result can be obtained considering the Maxwell fluxes in \eqref{RR sector lower rank fluxes}.}. As explained in   \cite{Lozano:2020txg, Lozano:2020sae}, they are the {\it restriction} of the fluxes to the internal space. Writing the Page fluxes in eqs.(\ref{fluxes23}) in terms of forms $J_i$ and ${\mathcal F}_i$ as,
 \begin{equation}
 \begin{split}
\hat{F}_1&=J_1,\qquad \hat{F}_3={\mathcal F}_{1}\wedge\text{vol}_{\text{AdS}_2}+	J_3,\qquad \hat{F}_5={\mathcal F}_{3}\wedge\text{vol}_{\text{AdS}_2}+J_5,\\
\hat{F}_7&={\mathcal F}_{5}\wedge\text{vol}_{\text{AdS}_2}+J_7,\qquad  \hat{F}_9={\mathcal F}_{7}\wedge\text{vol}_{\text{AdS}_2}.
 \end{split}
\end{equation}
The forms $J_i$ and ${\mathcal F}_i$ are,
\begin{equation}\label{forms}
\begin{split}
J_1 &= h'_8 \, \mathrm{d} \psi \, , \quad\quad
J_{3} = \frac{1}{4} \bigg(2h_8+\frac{u'(u\hat{h}_4'-\hat{h}_4u')}{2\hat{h}_4^2} \bigg) \text{vol}_{\text{S$^2$}}\wedge\text{d}\rho,\quad\quad  
J_{5} = - \hat{h}_4' \text{vol}_{\text{CY$_2$}} \wedge \mathrm{d} \psi \,\\
J_7&=- \frac{1}{4}\left( 2\hat{h}_4+\frac{u'(uh_8'-h_8u')}{2h_8^2} \right) \text{vol}_{\text{CY$_2$}}\wedge\text{vol}_{\text{S$^2$}}\wedge \text{d}\rho,\;\;\;\;
{\mathcal F}_1= \frac{1}{2} \left(h'_8(\rho-2\pi k)  - h_8\right)\text{d}\psi\, ,\\ 
{\mathcal F}_{3} &=\frac{1}{4}\left((\rho-2\pi k)h_8- \frac{\left(u - (\rho-2\pi k)  u' \right) ( u  \hat{h}_4' - \hat{h}_4  u'  )}{4 \hat{h}_4^2}\right)\text{vol}_{\text{S$^2$}} \wedge \mathrm{d} \rho, \\
{\mathcal F}_{5}&= \frac{1}{2} (\hat{h}_4 - (\rho-2\pi k) \hat{h}_4')\;  \text{vol}_{\text{CY$_2$}} \wedge \mathrm{d} \psi \, ,\\
{\mathcal F}_{7} &= -\frac{1}{4}\left((\rho-2\pi k) \hat{h}_4 -\frac{\left(u - (\rho-2\pi k)  u' \right) ( u  h_8' - h_8 u')}{4h_8^2}\right)  \text{vol}_{\text{CY$_2$}}\wedge\text{vol}_{\text{S$^2$}}\wedge \text{d}\rho.
\end{split}
\end{equation}
With the forms in eqs.\eqref{forms}, we construct the functional,
\begin{equation}\label{functionalcB}
\begin{split}
{\mathcal C} &= \int_{X_8} {\mathcal F}_{1}\wedge J_7+{\mathcal F}_{3}\wedge J_5-(J_1\wedge{\mathcal F}_{7}+J_3\wedge{\mathcal F}_{5})
\\
&=\int_{X_8}  \left(\frac{4\hat{h}_4 h_8-(u')^2}{8} -\frac{u^2}{16} \left( \frac{\hat{h}_4'^2}{\hat{h}_4^2} + \frac{h_8'^2}{h_8^2} \right) + \frac{u u'}{8} \left( \frac{\hat{h}_4'}{\hat{h}_4} + \frac{h_8'}{h_8} \right) \right) \text{vol}_{\text{CY}_2} \wedge  \text{vol}_{\text{S}^2}\wedge \text{d}\psi\wedge \text{d}\rho,
\end{split}
\end{equation}
We minimise the functional ${\mathcal C}$ by imposing the Euler-Lagrange equation for $u(\rho)$,
\begin{eqnarray}
	2u''=u\left(\frac{\hat{h}_4''}{\hat{h}_4}+\frac{h_8''}{h_8}\right).
\end{eqnarray}
This equation of motion is solved if,
 \begin{equation}
 \begin{split}
 	h_8''=0,\qquad \hat{h}_4''=0,\qquad u''=0,
 \end{split}
 \end{equation}
the first two are Bianchi identities for the background and the last is a BPS equation. 
%
The functional in eq.\eqref{functionalcB} can be rewritten as,
\begin{equation}
\begin{split}
{\mathcal C} &=\frac{1}{8}\int_{X_8}\left(4\hat{h}_4 h_8-(u')^2+\partial_\rho\left[\frac{u^2}{2}\left( \frac{\hat{h}_4'}{\hat{h}_4}+\frac{h_8'}{h_8} \right)\right]-\frac{u^2}{2}\left( \frac{\hat{h}_4''}{\hat{h}_4}+\frac{h_8''}{h_8} \right)
\right) \text{vol}_{\text{CY}_2} \wedge  \text{vol}_{\text{S}^2}\wedge \text{d}\psi\wedge \text{d}\rho.
\end{split}
\end{equation}
The last term (that would vanish in the absence of sources),  is proportional to the quotient of the number of flavours by the number of colours in each node. Using the condition that the flavours are sparse, as explained below eq.(\ref{nana}),  we see that its contribution is subleading in front  of the other terms. Furthermore, the boundary term gives a divergent contribution. Indeed, for the case $u=u_0$  and $\hat{h}_4, h_8 $ in eqs.\eqref{profileh4sp}-\eqref{profileh8sp} the boundary term reads,
\begin{equation}
\label{diver}
\begin{split}
\int_0^{2\pi (P+1)}	&\partial_\rho\left[\frac{u^2}{2}\left( \frac{\hat{h}_4'}{\hat{h}_4}+\frac{h_8'}{h_8} \right)\right]\text{vol}_{\text{CY}_2} \wedge  \text{vol}_{\text{S}^2}\wedge \text{d}\psi\wedge \text{d}\rho=-\lim_{\epsilon\to 0}\frac{2\pi u_0^2}{\epsilon}(\alpha_P+\mu_P+\beta_0+\nu_0)\text{Vol}_{\text{CY}_2}\\
&=-\lim_{\epsilon\to 0}\frac{2\pi u_0^2}{\epsilon}(Q_{\text{D}3}^{total}+Q_{\text{D}7}^{total})\text{Vol}_{\text{CY}_2},
\end{split}	
\end{equation}
  where we regularised  $\hat{h}_4(0)=h_8(0)=\hat{h}_4(2\pi(P+1))=h_8(2\pi(P+1))=\epsilon$. The divergence in eq.\eqref{diver} is associated with the presence of sources in the background as was found in \cite{Lozano:2020txg,Lozano:2020sae}.
  
   In summary,  the functional in eq.(\ref{functionalcB}) is proportional to the holographic central charge of eq.(\ref{hccharge-B}), plus a subleading contribution and a boundary term. For our infinite family of backgrounds, we have linked a calculation purely in terms of the NS-NS sector---eq.(\ref{hccharge-B}), with a calculation purely in terms of the Ramond-Ramond sector---eq.(\ref{sumdensities}), with the extremisation of a functional constructed as a restriction of the Ramond-Ramond forms to the internal space---eq.(\ref{functionalcB}). We believe that this may be a generic feature, worth exploring in backgrounds dual to various SCFTs in different dimensions.

\section{Conclusions}\label{conclusions}
 
 \paragraph{}We close this paper by presenting a short summary of the contents of this work and proposing future lines of investigation. 

This work presents two new infinite families of backgrounds with an AdS$_2$ factor.  The presentation focuses mostly on geometrical aspects of the new  solutions. The  new family of backgrounds in Section \ref{B-type} can be obtained by analytically continuing the backgrounds of  \cite{Lozano:2020txg} or via T-duality, on the Hopf-fibre of the S$^3$, from the solutions in \cite{Lozano:2020sae}. These connections are summarised in Figure \ref{diagrama}. A precise brane set-up was proposed for these backgrounds and the holographic central charge was calculated. We used the brane set-up to argue for a precise quiver. The  IR dynamics of such quivers should be the SCQMs dual to our backgrounds.

The family of AdS$_2$ backgrounds in Section \ref{B-type} and that in the paper \cite{Lozano:2020txg} have been shown to be connected to the solutions of \cite{Chiodaroli:2009yw, Chiodaroli:2009xh}. In fact, under certain circumstances they extend this class of solutions. The connection between these qualitatively different backgrounds requires of a subtle zoom-in procedure that we explained in detail in Section \ref{CGKsection}.

A second family of new backgrounds is presented in Section \ref{NATD}. These interesting solutions depend explicitly on two coordinates  (labelled as $\rho$ and $r$ in Section \ref{NATD}) and were obtained by the application of non-Abelian T-duality on the AdS$_3$ factor of the backgrounds in \cite{Lozano:2019emq}. We leave for future work to discuss the associated brane set-up, though it seems clear that the ideas described in \cite{Lozano:2016kum, Lozano:2016wrs, Lozano:2017ole, Itsios:2017cew, Lozano:2018pcp} will play an essential role in the global-definition of these solutions. By the same token, it would be interesting to study the integrability (or not) of the backgrounds presented here, as well as those in  \cite{Lozano:2020txg,Lozano:2020sae}. Integrable string backgrounds dual to field theories described by linear quivers in dimensions $d=2,4,6$, have been found in \cite{Filippas:2019ihy, Rigatos:2020igd, Nunez:2018qcj,Filippas:2019puw}. Similar techniques should probably apply for the $d=1$ case.

Finally,  in Section \ref{maxwellcentralhol} the holographic central charge defined in Section \ref{B-type}---a quantity computed solely in terms of the NS-NS sector of the backgrounds, has been connected with a calculation purely in terms of the Ramond-Ramond sector of our solutions. A functional whose extremisation yields the holographic central charge was also discussed. It should be interesting to find out if a similar structure occurs generically for other AdS$_{d+1}$ backgrounds.

\section*{Acknowledgements}  We would like to thank Niall Macpherson and Salomon Zacarias for very useful discussions.
\\
The work of CN is supported by STFC grant ST/T000813/1. Y.L. and A.R. are partially supported by the Spanish government grant
PGC2018-096894-B-100. AR is supported by CONACyT-Mexico.

\bibliography{Bibliography}

\providecommand{\href}[2]{#2}\begingroup\raggedright\begin{thebibliography}{10}

\bibitem{Lozano:2020txg}
Y.~Lozano, C.~Nunez, A.~Ramirez, and S.~Speziali, ``{New AdS$_2$ backgrounds
  and ${\cal N}=4$ Conformal Quantum Mechanics},''
  \href{http://arxiv.org/abs/2011.00005}{{\ttfamily arXiv:2011.00005
  [hep-th]}}.

\bibitem{Maldacena:1997re}
J.~M. Maldacena, ``{The Large N limit of superconformal field theories and
  supergravity},'' \href{http://dx.doi.org/10.1023/A:1026654312961,
  10.4310/ATMP.1998.v2.n2.a1}{{\em Int. J. Theor. Phys.} {\bfseries 38} (1999)
  1113--1133}, \href{http://arxiv.org/abs/hep-th/9711200}{{\ttfamily
  arXiv:hep-th/9711200 [hep-th]}}.
[Adv. Theor. Math. Phys.2,231(1998)].

\bibitem{Hanany:1996ie}
A.~Hanany and E.~Witten, ``{Type IIB superstrings, BPS monopoles, and
  three-dimensional gauge dynamics},''
  \href{http://dx.doi.org/10.1016/S0550-3213(97)00157-0}{{\em Nucl. Phys. B}
  {\bfseries 492} (1997) 152--190},
  \href{http://arxiv.org/abs/hep-th/9611230}{{\ttfamily arXiv:hep-th/9611230}}.

\bibitem{Apruzzi:2013yva}
F.~Apruzzi, M.~Fazzi, D.~Rosa, and A.~Tomasiello, ``{All AdS$_7$ solutions of
  type II supergravity},''
  \href{http://dx.doi.org/10.1007/JHEP04(2014)064}{{\em JHEP} {\bfseries 04}
  (2014) 064}, \href{http://arxiv.org/abs/1309.2949}{{\ttfamily arXiv:1309.2949
  [hep-th]}}.

\bibitem{Gaiotto:2014lca}
D.~Gaiotto and A.~Tomasiello, ``{Holography for (1,0) theories in six
  dimensions},'' \href{http://dx.doi.org/10.1007/JHEP12(2014)003}{{\em JHEP}
  {\bfseries 12} (2014) 003}, \href{http://arxiv.org/abs/1404.0711}{{\ttfamily
  arXiv:1404.0711 [hep-th]}}.

\bibitem{Cremonesi:2015bld}
S.~Cremonesi and A.~Tomasiello, ``{6d holographic anomaly match as a continuum
  limit},'' \href{http://dx.doi.org/10.1007/JHEP05(2016)031}{{\em JHEP}
  {\bfseries 05} (2016) 031}, \href{http://arxiv.org/abs/1512.02225}{{\ttfamily
  arXiv:1512.02225 [hep-th]}}.

\bibitem{Nunez:2018ags}
C.~Nunez, J.~M. Penin, D.~Roychowdhury, and J.~Van~Gorsel, ``{The
  non-Integrability of Strings in Massive Type IIA and their Holographic
  duals},'' \href{http://dx.doi.org/10.1007/JHEP06(2018)078}{{\em JHEP}
  {\bfseries 06} (2018) 078}, \href{http://arxiv.org/abs/1802.04269}{{\ttfamily
  arXiv:1802.04269 [hep-th]}}.

\bibitem{Brandhuber:1999np}
A.~Brandhuber and Y.~Oz, ``{The D-4 - D-8 brane system and five-dimensional
  fixed points},'' \href{http://dx.doi.org/10.1016/S0370-2693(99)00763-7}{{\em
  Phys. Lett. B} {\bfseries 460} (1999) 307--312},
  \href{http://arxiv.org/abs/hep-th/9905148}{{\ttfamily arXiv:hep-th/9905148}}.

\bibitem{Bergman:2012kr}
O.~Bergman and D.~Rodriguez-Gomez, ``{5d quivers and their AdS(6) duals},''
  \href{http://dx.doi.org/10.1007/JHEP07(2012)171}{{\em JHEP} {\bfseries 07}
  (2012) 171}, \href{http://arxiv.org/abs/1206.3503}{{\ttfamily arXiv:1206.3503
  [hep-th]}}.

\bibitem{Lozano:2012au}
Y.~Lozano, E.~\'O~Colg\'ain, D.~Rodriguez-G\'omez, and K.~Sfetsos,
  ``{Supersymmetric $AdS_6$ via T Duality},''
  \href{http://dx.doi.org/10.1103/PhysRevLett.110.231601}{{\em Phys. Rev.
  Lett.} {\bfseries 110} no.~23, (2013) 231601},
  \href{http://arxiv.org/abs/1212.1043}{{\ttfamily arXiv:1212.1043 [hep-th]}}.

\bibitem{DHoker:2016ujz}
E.~D'Hoker, M.~Gutperle, A.~Karch, and C.~F. Uhlemann, ``{Warped $AdS_6\times
  S^2$ in Type IIB supergravity I: Local solutions},''
  \href{http://dx.doi.org/10.1007/JHEP08(2016)046}{{\em JHEP} {\bfseries 08}
  (2016) 046}, \href{http://arxiv.org/abs/1606.01254}{{\ttfamily
  arXiv:1606.01254 [hep-th]}}.

\bibitem{DHoker:2016ysh}
E.~D'Hoker, M.~Gutperle, and C.~F. Uhlemann, ``{Holographic duals for
  five-dimensional superconformal quantum field theories},''
  \href{http://dx.doi.org/10.1103/PhysRevLett.118.101601}{{\em Phys. Rev.
  Lett.} {\bfseries 118} no.~10, (2017) 101601},
  \href{http://arxiv.org/abs/1611.09411}{{\ttfamily arXiv:1611.09411
  [hep-th]}}.

\bibitem{DHoker:2017zwj}
E.~D'Hoker, M.~Gutperle, and C.~F. Uhlemann, ``{Warped $AdS_6\times S^2$ in
  Type IIB supergravity III: Global solutions with seven-branes},''
  \href{http://dx.doi.org/10.1007/JHEP11(2017)200}{{\em JHEP} {\bfseries 11}
  (2017) 200}, \href{http://arxiv.org/abs/1706.00433}{{\ttfamily
  arXiv:1706.00433 [hep-th]}}.

\bibitem{Lozano:2018pcp}
Y.~Lozano, N.~T. Macpherson, and J.~Montero, ``{AdS$_{6}$ T-duals and type IIB
  AdS$_{6} \times$ S$^{2}$ geometries with 7-branes},''
  \href{http://dx.doi.org/10.1007/JHEP01(2019)116}{{\em JHEP} {\bfseries 01}
  (2019) 116}, \href{http://arxiv.org/abs/1810.08093}{{\ttfamily
  arXiv:1810.08093 [hep-th]}}.

\bibitem{Gaiotto:2009gz}
D.~Gaiotto and J.~Maldacena, ``{The Gravity duals of N=2 superconformal field
  theories},'' \href{http://dx.doi.org/10.1007/JHEP10(2012)189}{{\em JHEP}
  {\bfseries 10} (2012) 189}, \href{http://arxiv.org/abs/0904.4466}{{\ttfamily
  arXiv:0904.4466 [hep-th]}}.

\bibitem{ReidEdwards:2010qs}
R.~Reid-Edwards and j.~Stefanski, B., ``{On Type IIA geometries dual to N = 2
  SCFTs},'' \href{http://dx.doi.org/10.1016/j.nuclphysb.2011.04.002}{{\em Nucl.
  Phys. B} {\bfseries 849} (2011) 549--572},
  \href{http://arxiv.org/abs/1011.0216}{{\ttfamily arXiv:1011.0216 [hep-th]}}.

\bibitem{Aharony:2012tz}
O.~Aharony, L.~Berdichevsky, and M.~Berkooz, ``{4d N=2 superconformal linear
  quivers with type IIA duals},''
  \href{http://dx.doi.org/10.1007/JHEP08(2012)131}{{\em JHEP} {\bfseries 08}
  (2012) 131}, \href{http://arxiv.org/abs/1206.5916}{{\ttfamily arXiv:1206.5916
  [hep-th]}}.

\bibitem{Nunez:2019gbg}
C.~Nunez, D.~Roychowdhury, S.~Speziali, and S.~Zacarias, ``{Holographic aspects
  of four dimensional $\mathcal{N}=2$ SCFTs and their marginal deformations},''
  \href{http://dx.doi.org/10.1016/j.nuclphysb.2019.114617}{{\em Nucl. Phys. B}
  {\bfseries 943} (2019) 114617},
  \href{http://arxiv.org/abs/1901.02888}{{\ttfamily arXiv:1901.02888
  [hep-th]}}.

\bibitem{DHoker:2007zhm}
E.~D'Hoker, J.~Estes, and M.~Gutperle, ``{Exact half-BPS Type IIB interface
  solutions. I. Local solution and supersymmetric Janus},''
  \href{http://dx.doi.org/10.1088/1126-6708/2007/06/021}{{\em JHEP} {\bfseries
  06} (2007) 021}, \href{http://arxiv.org/abs/0705.0022}{{\ttfamily
  arXiv:0705.0022 [hep-th]}}.

\bibitem{DHoker:2008lup}
E.~D'Hoker, J.~Estes, M.~Gutperle, and D.~Krym, ``{Exact Half-BPS Flux
  Solutions in M-theory. I: Local Solutions},''
  \href{http://dx.doi.org/10.1088/1126-6708/2008/08/028}{{\em JHEP} {\bfseries
  08} (2008) 028}, \href{http://arxiv.org/abs/0806.0605}{{\ttfamily
  arXiv:0806.0605 [hep-th]}}.

\bibitem{Assel:2011xz}
B.~Assel, C.~Bachas, J.~Estes, and J.~Gomis, ``{Holographic Duals of D=3 N=4
  Superconformal Field Theories},''
  \href{http://dx.doi.org/10.1007/JHEP08(2011)087}{{\em JHEP} {\bfseries 08}
  (2011) 087}, \href{http://arxiv.org/abs/1106.4253}{{\ttfamily arXiv:1106.4253
  [hep-th]}}.

\bibitem{Couzens:2017way}
C.~Couzens, C.~Lawrie, D.~Martelli, S.~Schafer-Nameki, and J.-M. Wong,
  ``{F-theory and AdS$_{3}$/CFT$_{2}$},''
  \href{http://dx.doi.org/10.1007/JHEP08(2017)043}{{\em JHEP} {\bfseries 08}
  (2017) 043}, \href{http://arxiv.org/abs/1705.04679}{{\ttfamily
  arXiv:1705.04679 [hep-th]}}.

\bibitem{Lozano:2019emq}
Y.~Lozano, N.~T. Macpherson, C.~Nunez, and A.~Ramirez, ``{AdS$_3$ solutions in
  Massive IIA with small $\mathcal{N}=(4,0)$ supersymmetry},''
  \href{http://dx.doi.org/10.1007/JHEP01(2020)129}{{\em JHEP} {\bfseries 01}
  (2020) 129}, \href{http://arxiv.org/abs/1908.09851}{{\ttfamily
  arXiv:1908.09851 [hep-th]}}.

\bibitem{Lozano:2019jza}
Y.~Lozano, N.~T. Macpherson, C.~Nunez, and A.~Ramirez, ``{1/4 BPS solutions and
  the AdS$_3$/CFT$_2$ correspondence},''
  \href{http://dx.doi.org/10.1103/PhysRevD.101.026014}{{\em Phys. Rev. D}
  {\bfseries 101} no.~2, (2020) 026014},
  \href{http://arxiv.org/abs/1909.09636}{{\ttfamily arXiv:1909.09636
  [hep-th]}}.

\bibitem{Lozano:2019zvg}
Y.~Lozano, N.~T. Macpherson, C.~Nunez, and A.~Ramirez, ``{Two dimensional
  ${\cal N}=(0,4)$ quivers dual to AdS$_3$ solutions in massive IIA},''
  \href{http://dx.doi.org/10.1007/JHEP01(2020)140}{{\em JHEP} {\bfseries 01}
  (2020) 140}, \href{http://arxiv.org/abs/1909.10510}{{\ttfamily
  arXiv:1909.10510 [hep-th]}}.

\bibitem{Lozano:2019ywa}
Y.~Lozano, N.~T. Macpherson, C.~Nunez, and A.~Ramirez, ``{AdS$_3$ solutions in
  massive IIA, defect CFTs and T-duality},''
  \href{http://dx.doi.org/10.1007/JHEP12(2019)013}{{\em JHEP} {\bfseries 12}
  (2019) 013}, \href{http://arxiv.org/abs/1909.11669}{{\ttfamily
  arXiv:1909.11669 [hep-th]}}.

\bibitem{Lozano:2020bxo}
Y.~Lozano, C.~Nunez, A.~Ramirez, and S.~Speziali, ``{$M$-strings and AdS$_3$
  solutions to M-theory with small $\mathcal{N}=(0,4)$ supersymmetry},''
  \href{http://dx.doi.org/10.1007/JHEP08(2020)118}{{\em JHEP} {\bfseries 08}
  (2020) 118}, \href{http://arxiv.org/abs/2005.06561}{{\ttfamily
  arXiv:2005.06561 [hep-th]}}.

\bibitem{Faedo:2020nol}
F.~Faedo, Y.~Lozano, and N.~Petri, ``{Searching for surface defect CFTs within
  AdS$_3$},'' \href{http://dx.doi.org/10.1007/JHEP11(2020)052}{{\em JHEP}
  {\bfseries 11} (2020) 052}, \href{http://arxiv.org/abs/2007.16167}{{\ttfamily
  arXiv:2007.16167 [hep-th]}}.

\bibitem{Faedo:2020lyw}
F.~Faedo, Y.~Lozano, and N.~Petri, ``{New $\mathcal{N}=(0,4)$ AdS$_3$
  near-horizons in Type IIB},''
  \href{http://arxiv.org/abs/2012.07148}{{\ttfamily arXiv:2012.07148
  [hep-th]}}.

\bibitem{Dibitetto:2020bsh}
G.~Dibitetto and N.~Petri, ``{AdS$_3$ from M-branes at conical
  singularities},'' \href{http://arxiv.org/abs/2010.12323}{{\ttfamily
  arXiv:2010.12323 [hep-th]}}.

\bibitem{Dibitetto:2018gtk}
G.~Dibitetto and N.~Petri, ``{AdS$_{2}$ solutions and their massive IIA
  origin},'' \href{http://dx.doi.org/10.1007/JHEP05(2019)107}{{\em JHEP}
  {\bfseries 05} (2019) 107}, \href{http://arxiv.org/abs/1811.11572}{{\ttfamily
  arXiv:1811.11572 [hep-th]}}.

\bibitem{Gauntlett:2006ns}
J.~P. Gauntlett, N.~Kim, and D.~Waldram, ``{Supersymmetric AdS(3), AdS(2) and
  Bubble Solutions},''
  \href{http://dx.doi.org/10.1088/1126-6708/2007/04/005}{{\em JHEP} {\bfseries
  04} (2007) 005}, \href{http://arxiv.org/abs/hep-th/0612253}{{\ttfamily
  arXiv:hep-th/0612253}}.

\bibitem{Kim:2013xza}
N.~Kim, ``{Comments on $AdS_2$ solutions from M2-branes on complex curves and
  the backreacted K$\ddot{\text{a}}$hler geometry},''
  \href{http://dx.doi.org/10.1140/epjc/s10052-014-2778-6}{{\em Eur. Phys. J. C}
  {\bfseries 74} no.~2, (2014) 2778},
  \href{http://arxiv.org/abs/1311.7372}{{\ttfamily arXiv:1311.7372 [hep-th]}}.

\bibitem{Chiodaroli:2009yw}
M.~Chiodaroli, M.~Gutperle, and D.~Krym, ``{Half-BPS Solutions locally
  asymptotic to AdS(3) x S**3 and interface conformal field theories},''
  \href{http://dx.doi.org/10.1007/JHEP02(2010)066}{{\em JHEP} {\bfseries 02}
  (2010) 066}, \href{http://arxiv.org/abs/0910.0466}{{\ttfamily arXiv:0910.0466
  [hep-th]}}.

\bibitem{Chiodaroli:2009xh}
M.~Chiodaroli, E.~D'Hoker, and M.~Gutperle, ``{Open Worldsheets for Holographic
  Interfaces},'' \href{http://dx.doi.org/10.1007/JHEP03(2010)060}{{\em JHEP}
  {\bfseries 03} (2010) 060}, \href{http://arxiv.org/abs/0912.4679}{{\ttfamily
  arXiv:0912.4679 [hep-th]}}.

\bibitem{Corbino:2018fwb}
D.~Corbino, E.~D'Hoker, J.~Kaidi, and C.~F. Uhlemann, ``{Global half-BPS
  $AdS_2\times S^6$ solutions in Type IIB},''
  \href{http://dx.doi.org/10.1007/JHEP03(2019)039}{{\em JHEP} {\bfseries 03}
  (2019) 039}, \href{http://arxiv.org/abs/1812.10206}{{\ttfamily
  arXiv:1812.10206 [hep-th]}}.

\bibitem{Corbino:2020lzq}
D.~Corbino, ``{Warped $AdS_{2}$ and $SU(1,1|4)$ symmetry in Type IIB},''
  \href{http://arxiv.org/abs/2004.12613}{{\ttfamily arXiv:2004.12613
  [hep-th]}}.

\bibitem{Dibitetto:2019nyz}
G.~Dibitetto, Y.~Lozano, N.~Petri, and A.~Ramirez, ``{Holographic description
  of M-branes via AdS$_{2}$},''
  \href{http://dx.doi.org/10.1007/JHEP04(2020)037}{{\em JHEP} {\bfseries 04}
  (2020) 037}, \href{http://arxiv.org/abs/1912.09932}{{\ttfamily
  arXiv:1912.09932 [hep-th]}}.

\bibitem{Lozano:2020sae}
Y.~Lozano, C.~Nunez, A.~Ramirez, and S.~Speziali, ``{AdS$_2$ duals to ADHM
  quivers with Wilson lines},''
  \href{http://arxiv.org/abs/2011.13932}{{\ttfamily arXiv:2011.13932
  [hep-th]}}.

\bibitem{Lozano:2016kum}
Y.~Lozano and C.~N\'u\~nez, ``{Field theory aspects of non-Abelian T-duality
  and $ \mathcal{N} =$ 2 linear quivers},''
  \href{http://dx.doi.org/10.1007/JHEP05(2016)107}{{\em JHEP} {\bfseries 05}
  (2016) 107}, \href{http://arxiv.org/abs/1603.04440}{{\ttfamily
  arXiv:1603.04440 [hep-th]}}.

\bibitem{Lozano:2016wrs}
Y.~Lozano, N.~T. Macpherson, J.~Montero, and C.~Nunez, ``{Three-dimensional $
  \mathcal{N}=4 $ linear quivers and non-Abelian T-duals},''
  \href{http://dx.doi.org/10.1007/JHEP11(2016)133}{{\em JHEP} {\bfseries 11}
  (2016) 133}, \href{http://arxiv.org/abs/1609.09061}{{\ttfamily
  arXiv:1609.09061 [hep-th]}}.

\bibitem{Lozano:2017ole}
Y.~Lozano, C.~Nunez, and S.~Zacarias, ``{BMN Vacua, Superstars and Non-Abelian
  T-duality},'' \href{http://dx.doi.org/10.1007/JHEP09(2017)008}{{\em JHEP}
  {\bfseries 09} (2017) 008}, \href{http://arxiv.org/abs/1703.00417}{{\ttfamily
  arXiv:1703.00417 [hep-th]}}.

\bibitem{Itsios:2017cew}
G.~Itsios, Y.~Lozano, J.~Montero, and C.~Nunez, ``{The AdS$_{5}$ non-Abelian
  T-dual of Klebanov-Witten as a $ \mathcal{N}=1 $ linear quiver from
  M5-branes},'' \href{http://dx.doi.org/10.1007/JHEP09(2017)038}{{\em JHEP}
  {\bfseries 09} (2017) 038}, \href{http://arxiv.org/abs/1705.09661}{{\ttfamily
  arXiv:1705.09661 [hep-th]}}.

\bibitem{Macpherson:2014eza}
N.~T. Macpherson, C.~Nunez, L.~A. Pando~Zayas, V.~G.~J. Rodgers, and C.~A.
  Whiting, ``{Type IIB supergravity solutions with AdS$_{5}$ from Abelian and
  non-Abelian T dualities},''
  \href{http://dx.doi.org/10.1007/JHEP02(2015)040}{{\em JHEP} {\bfseries 02}
  (2015) 040}, \href{http://arxiv.org/abs/1410.2650}{{\ttfamily arXiv:1410.2650
  [hep-th]}}.

\bibitem{Bea:2015fja}
Y.~Bea, J.~D. Edelstein, G.~Itsios, K.~S. Kooner, C.~Nunez, D.~Schofield, and
  J.~A. Sierra-Garcia, ``{Compactifications of the Klebanov-Witten CFT and new
  AdS$_{3}$ backgrounds},''
  \href{http://dx.doi.org/10.1007/JHEP05(2015)062}{{\em JHEP} {\bfseries 05}
  (2015) 062}, \href{http://arxiv.org/abs/1503.07527}{{\ttfamily
  arXiv:1503.07527 [hep-th]}}.

\bibitem{Assel:2019iae}
B.~Assel and A.~Sciarappa, ``{On monopole bubbling contributions to
  \textquoteright{}t Hooft loops},''
  \href{http://dx.doi.org/10.1007/JHEP05(2019)180}{{\em JHEP} {\bfseries 05}
  (2019) 180}, \href{http://arxiv.org/abs/1903.00376}{{\ttfamily
  arXiv:1903.00376 [hep-th]}}.

\bibitem{DHoker:2007mci}
E.~D'Hoker, J.~Estes, and M.~Gutperle, ``{Gravity duals of half-BPS Wilson
  loops},'' \href{http://dx.doi.org/10.1088/1126-6708/2007/06/063}{{\em JHEP}
  {\bfseries 06} (2007) 063}, \href{http://arxiv.org/abs/0705.1004}{{\ttfamily
  arXiv:0705.1004 [hep-th]}}.

\bibitem{DHoker:2017mds}
E.~D'Hoker, M.~Gutperle, and C.~F. Uhlemann, ``{Warped $AdS_6\times S^2$ in
  Type IIB supergravity II: Global solutions and five-brane webs},''
  \href{http://dx.doi.org/10.1007/JHEP05(2017)131}{{\em JHEP} {\bfseries 05}
  (2017) 131}, \href{http://arxiv.org/abs/1703.08186}{{\ttfamily
  arXiv:1703.08186 [hep-th]}}.

\bibitem{Sfetsos:2010uq}
K.~Sfetsos and D.~C. Thompson, ``{On non-abelian T-dual geometries with Ramond
  fluxes},'' \href{http://dx.doi.org/10.1016/j.nuclphysb.2010.12.013}{{\em
  Nucl. Phys. B} {\bfseries 846} (2011) 21--42},
  \href{http://arxiv.org/abs/1012.1320}{{\ttfamily arXiv:1012.1320 [hep-th]}}.

\bibitem{Lozano:2011kb}
Y.~Lozano, E.~O~Colgain, K.~Sfetsos, and D.~C. Thompson, ``{Non-abelian
  T-duality, Ramond Fields and Coset Geometries},''
  \href{http://dx.doi.org/10.1007/JHEP06(2011)106}{{\em JHEP} {\bfseries 06}
  (2011) 106}, \href{http://arxiv.org/abs/1104.5196}{{\ttfamily arXiv:1104.5196
  [hep-th]}}.

\bibitem{Itsios:2012dc}
G.~Itsios, Y.~Lozano, E.~O~Colgain, and K.~Sfetsos, ``{Non-Abelian T-duality
  and consistent truncations in type-II supergravity},''
  \href{http://dx.doi.org/10.1007/JHEP08(2012)132}{{\em JHEP} {\bfseries 08}
  (2012) 132}, \href{http://arxiv.org/abs/1205.2274}{{\ttfamily arXiv:1205.2274
  [hep-th]}}.

\bibitem{Itsios:2012zv}
G.~Itsios, C.~Nunez, K.~Sfetsos, and D.~C. Thompson, ``{On Non-Abelian
  T-Duality and new N=1 backgrounds},''
  \href{http://dx.doi.org/10.1016/j.physletb.2013.03.033}{{\em Phys. Lett. B}
  {\bfseries 721} (2013) 342--346},
  \href{http://arxiv.org/abs/1212.4840}{{\ttfamily arXiv:1212.4840 [hep-th]}}.

\bibitem{Itsios:2013wd}
G.~Itsios, C.~Nunez, K.~Sfetsos, and D.~C. Thompson, ``{Non-Abelian T-duality
  and the AdS/CFT correspondence:new N=1 backgrounds},''
  \href{http://dx.doi.org/10.1016/j.nuclphysb.2013.04.004}{{\em Nucl. Phys. B}
  {\bfseries 873} (2013) 1--64},
  \href{http://arxiv.org/abs/1301.6755}{{\ttfamily arXiv:1301.6755 [hep-th]}}.

\bibitem{Alvarez:1993qi}
E.~Alvarez, L.~Alvarez-Gaume, J.~Barbon, and Y.~Lozano, ``{Some global aspects
  of duality in string theory},''
  \href{http://dx.doi.org/10.1016/0550-3213(94)90067-1}{{\em Nucl. Phys. B}
  {\bfseries 415} (1994) 71--100},
  \href{http://arxiv.org/abs/hep-th/9309039}{{\ttfamily arXiv:hep-th/9309039}}.

\bibitem{Couzens:2018wnk}
C.~Couzens, J.~P. Gauntlett, D.~Martelli, and J.~Sparks, ``{A geometric dual of
  $c$-extremization},'' \href{http://dx.doi.org/10.1007/JHEP01(2019)212}{{\em
  JHEP} {\bfseries 01} (2019) 212},
  \href{http://arxiv.org/abs/1810.11026}{{\ttfamily arXiv:1810.11026
  [hep-th]}}.

\bibitem{Gauntlett:2018dpc}
J.~P. Gauntlett, D.~Martelli, and J.~Sparks, ``{Toric geometry and the dual of
  $c$-extremization},'' \href{http://dx.doi.org/10.1007/JHEP01(2019)204}{{\em
  JHEP} {\bfseries 01} (2019) 204},
  \href{http://arxiv.org/abs/1812.05597}{{\ttfamily arXiv:1812.05597
  [hep-th]}}.

\bibitem{Filippas:2019ihy}
K.~Filippas, ``{Non-integrability on AdS$_{3}$ supergravity backgrounds},''
  \href{http://dx.doi.org/10.1007/JHEP02(2020)027}{{\em JHEP} {\bfseries 02}
  (2020) 027}, \href{http://arxiv.org/abs/1910.12981}{{\ttfamily
  arXiv:1910.12981 [hep-th]}}.

\bibitem{Rigatos:2020igd}
K.~S. Rigatos, ``{Non-integrability in AdS$_3$ vacua},''
  \href{http://arxiv.org/abs/2011.08224}{{\ttfamily arXiv:2011.08224
  [hep-th]}}.

\bibitem{Nunez:2018qcj}
C.~Nunez, D.~Roychowdhury, and D.~C. Thompson, ``{Integrability and
  non-integrability in $ \mathcal{N}=2 $ SCFTs and their holographic
  backgrounds},'' \href{http://dx.doi.org/10.1007/JHEP07(2018)044}{{\em JHEP}
  {\bfseries 07} (2018) 044}, \href{http://arxiv.org/abs/1804.08621}{{\ttfamily
  arXiv:1804.08621 [hep-th]}}.

\bibitem{Filippas:2019puw}
K.~Filippas, C.~Nunez, and J.~Van~Gorsel, ``{Integrability and holographic
  aspects of six-dimensional $ \mathcal{N}=\left(1,\ 0\right) $ superconformal
  field theories},'' \href{http://dx.doi.org/10.1007/JHEP06(2019)069}{{\em
  JHEP} {\bfseries 06} (2019) 069},
  \href{http://arxiv.org/abs/1901.08598}{{\ttfamily arXiv:1901.08598
  [hep-th]}}.

\end{thebibliography}\endgroup

\end{document}